# Tuning the Coercive Field in Ferroelectric $Hf_{0.5}Zr_{0.5}O_2$-$Al_2O_3$ Heterostructures via Interfacial Charge Dynamics

Marshall B. Frye[1], Chanyoung Kim[1], Jeong-Woo Sun[1], John Wellington-Johnson[1], Lance Fernandes[2], Prasanna Venkatesan Ravindran[2], Bogdan Dryzhakov[3], TaeYoung Song[2], Mengkun Tian[4], Asif I. Khan[1,2], Lauren M. Garten[1]*

[1] School of Materials Science and Engineering, Georgia Institute of Technology, Atlanta, Georgia 30332, United States

[2] School of Electrical and Computer Engineering, Georgia Institute of Technology, Atlanta, Georgia 30332, United States

[3] Center for Nanophase Materials Sciences, Oak Ridge National Laboratory, Oak Ridge, Tennessee 37831, United States

[4] The Institute for Matter and Systems, Georgia Institute of Technology, Atlanta, Georgia 30332, United States

*Corresponding Author: Lauren.garten@mse.gatech.edu

**Abstract**

Interleaving dielectric layers into ferroelectric $Hf_{0.5}Zr_{0.5}O_2$ (HZO) films increases the memory window (MW) beyond what is accounted for by the dielectric constants. Determining the physical mechanisms behind these MW improvements is critical to reaching the full potential of HZO FeNAND. Here, we show that MW improvements stem from the interfacial charge dynamics enabled by oxygen vacancies at the interlayer interface. X-ray photoelectron spectroscopy(XPS) etching experiments demonstrate increased off-stoichiometry at the interface, with a 2.3x increase in oxygen vacancies. Polarization-dependent XPS and first-order reversal curves(FORC) show that tunneling between interfacial defect states causes a bidirectional internal bias of 0.56MV/cm. The impact of defects is further corroborated through phase-field modeling(PFM), which only recreates the coercive fields, FORC, and internal bias for defect densities and tunneling barrier heights that are consistent with experiment, quantitatively capturing an internal electric field of 0.55 MV/cm. The phase field models are then used to simulate 36 devices with varied charge densities and dielectric thickness to provide a predictive framework for further improvements in the MW of interlayer HZO. These findings redefine the role of defects in ferroelectric HZO from deleterious to engineerable and provide critical insights into how to tune ferroelectric device architectures for improved memory.

## 1. Introduction

Reducing the energy consumption of nonvolatile memory while simultaneously increasing memory density is critically needed to surpass the memory bottlenecks currently constraining computational scaling. Charge trap flash (CTF) NAND is the current standard for nonvolatile memory, but high operating voltages (>20 V), slow switching speeds (~$10^{-3}$ s), and cell-to-cell interference limit device performance and scaling.[1] $Hf_{0.5}Zr_{0.5}O_2$ (HZO) based ferroelectric NAND (FeNAND) is predicted to surpass these limitations with a 96% decrease in energy consumption,[2] and a 1000x increase in switching speed.[3] However, further increases in the memory window (MW) of HZO FeNAND are required to achieve quad-level cell (QLC) operation while allowing scaling to thicknesses below 20 nm.[1,4] The memory window in FeNAND is related to the ferroelectric response through the following equation: $MW = \Delta V_{th} \propto 2 \cdot E_c \cdot t$, where $E_c$ is the ferroelectric coercive field, and t is the ferroelectric thickness.[3] Since the coercive field of a ferroelectric is typically a fixed property, alternative approaches to increase the memory window are needed to keep pace with the downscaling of the film thickness.

Two primary approaches have emerged to increase the memory window in FeNAND: introducing a gate-side dielectric[2,5,6] or embedding a dielectric layer within the HZO.[7] In gate-side dielectric structures, the memory-window expansion is largely attributed to charge injection on the gate-side which reduces the retention of the polarization state over time.[1,2,8–11] In contrast, inserting the dielectric between two ferroelectric layers (gate–ferroelectric–dielectric–ferroelectric–channel) enables substantial memory window enhancement without compromising retention.[1,2,8–11] Embedding the ferroelectric-dielectric (F-D) interfaces has enabled record-breaking retention and energy efficiency in $Hf_{0.5}Zr_{0.5}O_2$.[12] Improvements in memory window and retention have also been seen in previous literature which was attributed to interfacial charges with the addition of dielectric layer,[11,15] but the mechanisms was not directly assessed. Still, the increase in memory window cannot be accounted for solely by the addition of a dielectric in series, raising the question: what is the underlying physical mechanism that is driving these improvements and how can these contributions be tuned to control the memory window?

Fluidic imprint –the variable and history-dependent asymmetric shift in ferroelectric coercive field[13,14] – has previously been proposed as a mechanism to increase the memory window by dynamically shifting the coercive fields with cycling in HZO.[15] Fluidic imprint has been attributed to charge trapping at defect sites at electrode interfaces.[13,14,16,17] Additionally, the stability of the ferroelectric phase and the polarization switching of HZO are known to be impacted by the presence of charged defects, particularly by oxygen vacancies.[18,19] Although fluidic imprint has been proposed as a mechanism to tune device performance in HZO, a definitive link between charged defects and ferroelectric performance has yet to be established, and competing mechanisms remain under debate.[1,2,20–23] In particular, the impact of polarization on the formation, redistribution, and electrostatic coupling of charged defects in embedded interlayer devices has not been resolved. Developing a mechanistic understanding is crucial because memory-window enhancement is not an isolated performance metric, but a fingerprint of the electrostatic and switching physics that ultimately define ferroelectric device reliability and scalability. Therefore, understanding the impact of defects and determining how to use these defects to control the coercive field in HZO is central to enabling high memory density FeNAND.

In this work, we provide a direct connection between charged defects and ferroelectric device performance. We first identify the critical defects and then demonstrate that under sufficiently large applied electric fields, the redistribution of electrons across a dielectric interlayer changes the occupancy of interfacial defects, reversing the internal field, fundamentally altering the switching process and increasing the coercive field. Spectroscopic and electrical signatures of polarization-dependent charge redistribution confirm a reversible internal field arising from electron redistribution at the interface only for devices with a dielectric interlayer. The internal fields lead to a threefold increase in coercive field with the insertion of a dielectric. Using the measured polarization-electric field dependence, defect concentrations, and band structure, a phase-field model is developed to track how polarization switching is impacted by dielectric interlayers. The model reveals two distinct switching pathways–interfacial charge-mediated switching and polarization-driven charge tunneling–depending on dielectric thickness, material, and interfacial defect density. This model clearly shows that polarization reversal is not dictated solely by ferroelectric energetics, but also by the coupled electrostatics of defects and electronic compensation at the interface. These findings establish defect-mediated interfacial charge dynamics as a route to engineer the coercive field to improve ferroelectric memory.

## 2. Results

### 2.1. Increased coercive fields with interlayer insertion in HZO

Understanding the physical mechanisms driving the increase in memory window of embedded dielectric layer devices begins with creating the ferroelectric-dielectric interface. Here, two device architectures are compared:

1) A continuous 19 nm HZO metal-ferroelectric-metal device used as a control, labeled as 19H.
2) A metal-ferroelectric-insulator-ferroelectric-metal device with an $Al_2O_3$ interlayer inserted between two 8 nm layers of HZO. The interlayer thickness is 3 nm unless otherwise specified and labeled as 8H-3A-8H.

$Al_2O_3$ is used as the dielectric interlayer material because it can be processed by ALD under conditions that are compatible with the HZO film fabrication and has a breakdown field >10 MV/cm.[24] Scanning transmission electron micrographs (STEM) of both the 19H and 8H-3A-8H device types are shown in **Figure 1a**, and **1b,** respectively. The STEM images reveal polycrystalline HZO with clear interfaces between the $Al_2O_3$, HZO, and tungsten electrode layers. The $Al_2O_3$ interlayer thickness is approximately 3 nm. Grazing-incidence X-ray diffraction (GIXRD) taken on both device architectures, displayed in **Figure S1,** shows that the primary phase for both is the ferroelectric $Pca2_1$ orthorhombic phase of HZO.

Comparing the ferroelectric response of the control devices to the interlayer devices shows how inserting dielectric interlayers modulates ferroelectric device performance. The polarization-electric field (P(E)) hysteresis loops for representative interlayer and control devices are shown in **Figure 1c**. Both device types exhibit characteristic features of ferroelectric hysteresis, including non-zero remanent polarization, relatively sharp tips with a positive slope upon approaching polarization saturation, a sharp change in curvature in the hysteresis in the second and fourth quadrants, and a well-defined coercive field.[25] The remanent polarizations (taken from the y-intercept of the P(E) loops) of the two device types are comparable, with an average $2P_r$ of 27.5 μC/cm$^2$ for 19H control and an average $2P_r$ of 24.7 μC/cm$^2$ for 8H-3A-8H interlayer devices. The ferroelectric hysteresis observed in the P(E) measurements is further corroborated by the electric field-dependent current density (J(E)) for both device types, presented in **Figure 1d** because the maximum in the current density exhibits peaks near the coercive fields measured in the P(E) loops for both device architectures. Additionally, polarization-up-negative-down (PUND) measurements are also provided to further support the determination of the remanent polarization. **Figure S2** shows the voltage waveform and resulting current for a PUND measurement on 8H-3A-8H sample. The extracted positive and negative remanent polarization are 22.35 μC/cm$^2$ and 25.23 μC/cm$^2$, respectively, consistent with the P(E) results.

The most notable difference between the ferroelectric response of the two device architectures is the increase in coercive field ($E_c$) with interlayer insertion. The average coercive field of 19H control devices is 0.96 $\pm$ 0.01 MV/cm, while that of the 8H-3A-8H interlayer devices is 3.10 $\pm$ 0.04 MV/cm. The coercive field of 19H is in the range of previously reported values for HZO deposited by atomic layer deposition,[26,27] whereas 8H-3A-8H has a 3.22 times increase in the coercive field. Since the coercive field of HZO is known to vary with thickness, an 8 nm thick continuous HZO device is also measured as a control to assess the expected coercive field for each HZO layer within the interlayer devices. **Figure S3** shows P(E) loops for an 8 nm thick HZO with, with a coercive field of +1.65 MV/cm and –1.29 MV/cm and remanent polarization of +21.05 μC/cm$^2$ and –22.16 μC/cm$^2$. So, the difference in coercive field cannot be accounted for by the difference in film thickness. The increase in coercive field cannot be accounted for by the decrease in HZO thickness or by the addition of a lower permittivity layer in series as that would only account for a 0.8 MV/cm difference (calculation description provided in **Supplemental Note 1**). Furthermore, it is unlikely that the change stems from the incorporation of aluminum into HZO from the $Al_2O_3$ interlayer, as the incorporation of aluminum has been shown to decrease rather than increase the coercive field.[28,29] Furthermore, prior research has shown that $Al_2O_3$ layers thicker than 0.2 nm do not diffuse appreciably into HZO.[30]

To understand the root-cause of the distinctive coercive field differences of the interlayer devices, the measured P(E) hysteresis of both device architectures is compared to P(E) loops simulated using a time-dependent Landau-Ginzburg-Devonshire (LGD) phase-field model. The phase-field modeling is performed through the FERRET module[31] based on the Multiphysics Object Oriented Simulation Environment.[32,33] Further information on the phase-field modeling procedure can be found in the **Methods** section. The simulated hysteresis of 19H (**Figure 1e**) is in strong agreement with the experimental results with a root mean square error, RMSE = 1.0, indicating that the model accurately captures the switching behavior of the control device.

For the modeling of the interlayer device, shown in **Figure 1f**, the simulation incorporates the dielectric and elastic constants of the $Al_2O_3$ interlayer. Even when accounting for the electric field drop across the $Al_2O_3$ interlayer and the disparate elastic constants, the simulation fails to reproduce the large coercive field seen in the measured P(E) loops of the interlayer devices. The model underestimates the average coercive field by 42.7%, demonstrating that elastic and dielectric contributions alone cannot account for the observed modulation of ferroelectric switching. The phase-field modeling indicates that there must be additional contributions that impact the polarization switching in the interlayer devices. The increase in coercive field could be indicative of fluidic imprint, which has been attributed to charge trapping at defect sites.[13,14] So, the next steps are to investigate the presence and impact of defects in the HZO interlayer devices.

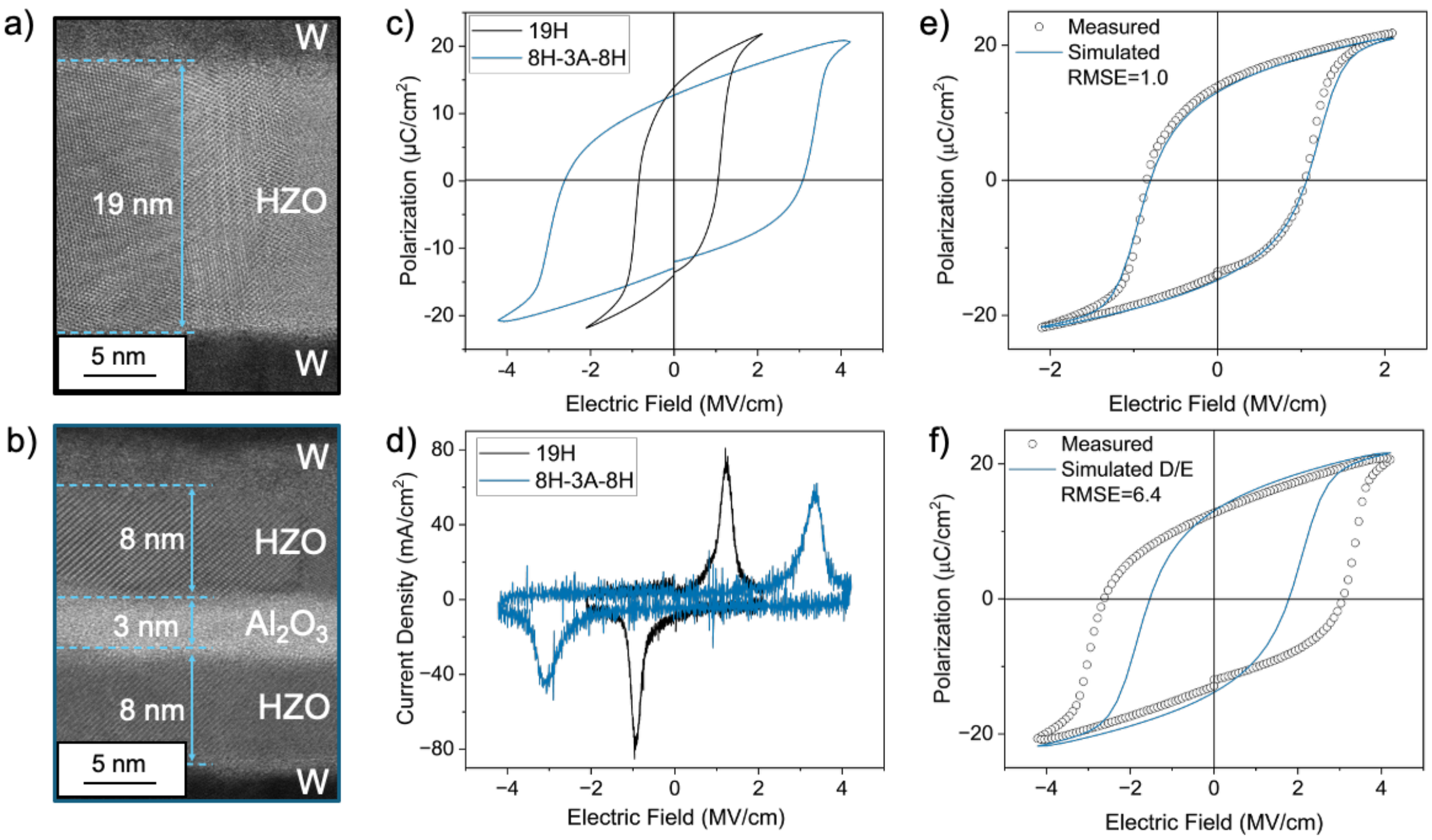

**Figure 1. Interlayer engineering modifies ferroelectric switching in HZO devices. 1a.** Scanning transmission electron micrograph of a control device (19H); **1b.** Scanning transmission electron micrograph of an interlayer device (8H-3A-8H), showing the $Al_2O_3$ interlayer; **1c.** Polarization-electric field (P(E)) hysteresis of representative 19H HZO thin film; **1d.** P(E) hysteresis of an 8H-3A-8H interlayer device; **1e.** Comparison between the phase-field model of the P(E) hysteresis and the experimental results for a 19H control device; and for an 8H-3A-8H interlayer device in **1f**.

### 2.2. Determining the presence and impact of defects at the ferroelectric-dielectric interface

The next step towards understanding the mechanisms contributing to the increased coercive field in the interlayer devices is to measure the interfacial defect states that form at the interlayer. X-ray photoelectron spectroscopy (XPS) provides information on the stoichiometry of the films and the defects at the ferroelectric-dielectric (F-D) interface.[34,35] Representative XPS core-level measurements and fits of the Hf 4f, Zr 3d, Al 2p, and O1s peaks are shown in **Figure**

**2a-d**, respectively. Initial measurements are taken on the bare surface of the film for each device type. After these initial measurements, the films are subsequently etched with an argon cluster tool to the top of the F-D interface (or an equivalent time for the 19H control devices). Scans are taken periodically to assess the etch depth as shown by the depth profile in **Figure S5**. The etching progresses until the calculated thickness of the HZO layer on top of the $Al_2O_3$ is 3.5 nm (the thickness calculations are described further in the methods section). The 19H control devices etched under the same conditions showed negligible changes in stoichiometry, indicating that the etching process does not damage HZO. Thus, the argon cluster is used for all subsequent etching because of its minimal damage compared to monatomic argon etching.[35] The average cation ratios extracted from XPS are 1:1.02 (Zr:Hf) for 19H films and 1:1.05 for 8H-3A-8H films, confirming the samples are close to the targeted 1:1 composition. The cation composition and oxidation state ratios before and after etching are shown in **Table 1**.

**Table 1**. Composition for the 8H-3A-8H and 19H samples determined by XPS peak fitting before and after etching to the ferroelectric-dielectric interface in the 8H-3A-8H or for an equivalent time for the 19H films.

| | **8H-3A-8H** | | **19H** | |
|---|---|---|---|---|
| **Element** | **Surface** | **Interface** | **Surface** | **Bulk** |
| Hf | 50.9% | 35.2% | 50.6% | 49.1% |
| Zr | 49.1% | 33.4% | 49.4% | 50.9% |
| Al | 0% | 31.4% | 0% | 0% |
| $Hf^{3+}$/Hf | 4.1% | 15.0% | 5.7% | 5.7% |
| $Zr^{3+}$/Zr | 5.9% | 8.4% | 3.4% | 4.6% |

In HZO, oxygen vacancies are typically inferred from $Hf^{3+}$ and $Zr^{3+}$ components of the Hf 4f and Zr 3d spectra, indicative of locally reduced cation coordination. Each cation with a 3+ oxidation state corresponds to half of an oxygen vacancy to maintain charge neutrality (i.e., $Hf^{3+}O^{2-}_{1.5}$).[34,36] The $3^+$/total cation ratios are given in **Table 1**, which were used to calculate the oxygen vacancy concentration using the method from reference: [34]. Following etching, oxygen vacancy concentration increases from 1.14% to 1.29% in the 19H control devices and from 1.25% to 2.93% in the 8H-3A-8H interlayer devices. The composition at the surface of all samples lies within the expected 1-2% oxygen vacancy concentrations that are commonly reported for bulk ferroelectric HZO.[37] But the 2.34x increase in defect concentration at the interlayer interface in the 8H-3A-8H samples indicates an increase in the oxygen vacancies localized at the interface compared to bulk HZO.

Photoluminescence (PL) and cathodoluminescence (CL) provide further insight into the defects present in each device type. The precise energy level of the oxygen vacancy defect states is not uniquely defined, as reported energies typically range from 0.8 to 2 eV below the HZO conduction band, including both shallow and deep defect levels.[38] Furthermore, defects at dielectric-HZO interfaces are reported to be further below the conduction band than in bulk HZO.[39] Photoluminescence of the 8H-3A-8H devices, shown in **Figure S6a,** is skewed to higher energy compared to 19H. Emission in the blue light energy range from HZO involves deep defect states,[40,41] indicating a greater relative contribution from deep defect states at the F-D interface. Cathodoluminescence measured at 10 K, exhibits emission consistent with transitions from shallow traps to deep defect states. As shown in **Figure S6b**, the CL from interlayer devices shows reduced broadening which suggests a narrower distribution of emissive defect states in the interlayer devices. Along with the higher spectral weight at high-energy transitions observed in PL, these findings indicate that the interlayer promotes a higher density of deep defect states with a more energetically uniform character, consistent with reduced spatial inhomogeneity in defect concentration at the interlayer interfaces.

XPS valence band spectroscopy builds on the PL and CL results to provide further context into electronically occupied defect states.[35,42–44] Representative valence spectra of 19H and 8H-3A-8H devices before and after etching are shown in **Figure S7a** and **Figure S7b**, respectively. In both cases, the shape of the valence band is consistent with prior reports of the O 2p band.[35,44,45] Occupied defect states, labelled as midgap states, are observed in the valence spectra at binding energies lower than the valence band edge. These midgap states are characteristic of electronic states that have previously been attributed to oxygen vacancies from density functional theory calculations.[35,44,45] For the 19H valence spectra, the valence band and midgap state intensity show little change before and after etching, as highlighted in the zoomed in region in **Figure 2e**. In contrast, the interlayer device exhibits a pronounced increase in intensity of the midgap states after etching, shown in **Figure 2f**. The increase in midgap states at the F-D interface is consistent with the increase in the off-stoichiometry seen in the core-level spectra at the interlayer interface. The increased density of electronically compensated defects localized near the interface is consistent with the defect-mediated contributions that are proposed to increase the coercive field in the interlayer devices.

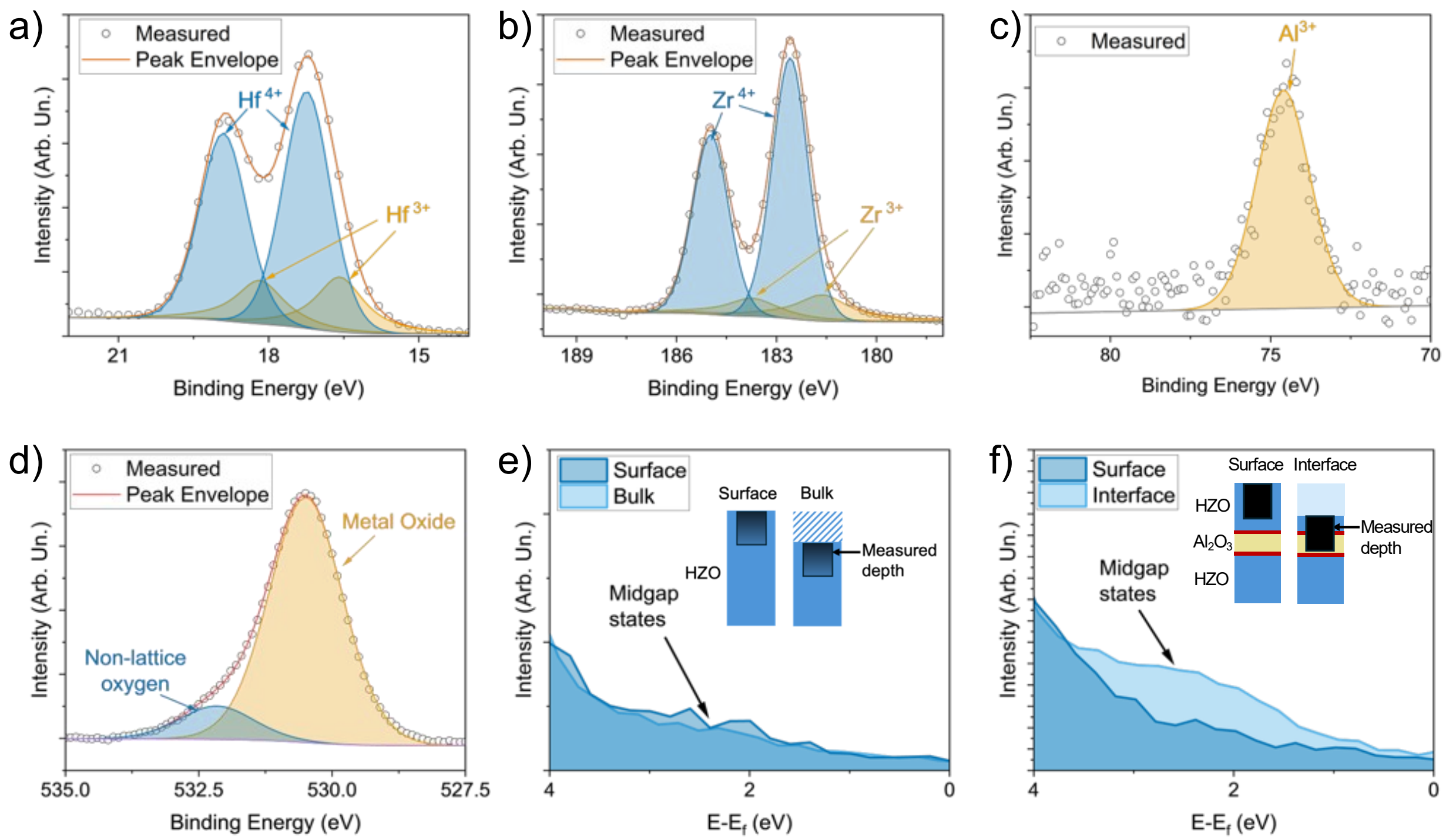

**Figure 2**. **XPS defect states and valence spectra before and after etching**. Representative measured XPS peaks and peak fitting for the 8H-3A-8H device after etching: **2a.** Hf 4f; **2b.** Zr 3d; **2c.** Al 2p; and **2d.** O 1s peaks. **2e.** XPS valence band spectra for representative control sample before and after etching **2f.** XPS valence band spectra for an interlayer sample before and after etching.

### 2.3. Measuring the band alignments in interlayer devices

The next step towards building a model to determine the impact of defects on ferroelectric switching requires determining the band alignment between layers because the probability of carriers tunneling across the interlayer into these defect traps is dependent upon the band alignments. The band structure of each layer is determined using a combination of electronic and photonic spectroscopic characterization techniques. The valence band spectra (**Figure S8a**) are used to identify both the valence band minima and fermi energy. The measured Fermi energy is then used along with the secondary electron cutoff spectra (**Figure S8b**) to determine the work function of HZO. Finally, the direct band gap was measured using reflection electron energy loss spectroscopy, shown in **Figure S8c**. The same techniques are applied to measure the band diagram for $Al_2O_3$ (**Figure S8e-g**). The measured band diagrams of the HZO and $Al_2O_3$ layers are shown in **Figure S8d** and **S8h**, respectively. Further details on fitting and measurements can be found in the **Methods** and in reference [46]. Taken collectively these band structure measurements provide the inputs needed to model the charge carrier dynamics across the interlayer.

### 2.4 Developing a phase-field model to assess the impact of interfacial defects on ferroelectric switching

Measuring the interfacial vacancies and the band alignments provides the inputs needed to model the carrier dynamics during ferroelectric switching in HZO. Because it has previously be established that there is minimal oxygen vacancy diffusion at room temperature -the diffusion coefficient of oxygen in ferroelectric HZO is negligible at room temperature (1.6 x $10^{-26}$ $cm^2/s$) and elevated temperature (4 – 7 x $10^{-17}$ $cm^2/s$ at 400 °C),[47] - the oxygen vacancies and all ionic defects are treated as stationary in the phase-field model. The interlayer device is modeled with a fixed concentration of positively charged oxygen vacancies on either side of the $Al_2O_3$ with an equal concentration of mobile electrons capable of tunneling across the dielectric to emulate the increased defects observed at the interface by XPS.[16,48,49] The initial concentration of defects at the interface was set to 0.6 $nm^{-2}$, in accordance with prior reports from references [50,51] and within the range of defects from the XPS findings. Tunneling is used in the model as the dominant charge transport mechanism across the interlayer as it has previously been established in prior literature.[15,48] While there are likely other mechanisms of conduction which are at play in these devices, the impact of tunneling in the interlayer devices is clearly demonstrated in the thickness dependence of the P(E) loops in **Figure S4a-c**. For films with a dielectric layer thickness greater than 1 nm, there is a near exponential decrease in polarization. Additionally, while the increase in coercive field starts to level off as the interlayer thickness increases, there is an increase in the imprint, which is consistent with a reduction in switched charge with increased barrier thickness. Therefore, initial modeling focuses on tunneling from fixed charge layers on either side of the interlayer interfaces.

The P(E) hysteresis of the 8H-3A-8H stack is simulated across a range of defect energies from 1.2 -1.8 eV, corresponding to tunneling barriers ($\varphi_b$) from 1.8 - 2.4 eV consistent with the barrier heights determined from the measured XPS and REELS results. **Figure S9a** shows the polarization-electric field loops calculated using the phase-field model for an interlayer device with a tunneling barrier height of $\varphi_b$=1.8 eV (corresponding to a defect energy of 1.2 eV below the conduction band). At room temperature, the calculated tunneling rate exceeded the expected Schottky emission across the interface by several orders of magnitude, in line with a recent report of electron tunneling through an $Al_2O_3$ layer from $HfO_2$.[48] The simulated hysteresis deviates from experiment, with an overestimated polarization and an underestimated coercive field. The mismatch between experimental and simulated hysteresis loops results in a root mean square error (RMSE) of 5.4. The misestimates are attributed to using too low of a barrier which causes premature electron tunneling, switching the direction of the internal field at reduced applied bias, inducing polarization switching. Furthermore, a high proportion of electrons tunnel, resulting in a large internal bias field that overestimates the polarization. Increasing the barrier height further to 2.4 eV (**Figure S9c**) reduces the electron transfer, suppressing the internal field, resulting in an underestimated coercive field and an increase in RMSE to 3.3. When the tunneling barrier is too high, carriers cannot readily traverse the interlayer making electronic compensation less effective and the interlayer defects no longer substantially modulate switching. (Note, while the distribution of electrons and positively charged oxygen vacancies was initially set to be equal on both F-D interfaces, any asymmetry that may form during deposition would lead to an imprint if the tunnel barrier is too high.)

Overall, the best fit to experimental P(E) is for a tunneling barrier of 2.1 eV (corresponding to a defect energy 1.5 eV below the conduction band). Setting the tunneling barrier to 2.1 eV (**Figure S9b**) results in a significant improvement in agreement between measured and simulated P(E) hysteresis, with an RMSE of 3.1. Importantly, the average coercive field is simulated within 4% of the experimental values, compared to the 42.7% error of the simulations without defects. In all cases, the simulated polarization of the 8H-3A-8H stack slightly overestimates experimental results. A general decrease in polarization in the 8H-3A-8H devices compared to the continuous films is anticipated due to depolarization stemming from the interfacial charge layers, which is already accounted for in the electrostatics and dynamics of the model. One potential explanation for the additional reduction in polarization between the modeled and experimental data could stem from a difference in the amount of the ferroelectric phase in the interlayer devices. X-ray diffraction (**Figure S1**) shows a decrease in the nonferroelectric monoclinic phase of HZO with interlayer insertion. However, it is not possible to definitively deconvolute the polar orthorhombic phase from the nonpolar tetragonal and orthorhombic phases due to the Scherrer broadening of the XRD peaks.[47] A reduction in remanent polarization was observed for 5 nm thick HZO compared to the 19 nm thick HZO control, shown in **Figure S9d**, indicating that the decrease is intrinsic to reduced HZO thickness and not from the $Al_2O_3$. Still, the reduced polarization in thinner HZO films indicates the difference between devices likely stems predominantly from a decrease in the ferroelectric phase. Accordingly, the 8H-3A-8H devices were simulated with a barrier height of 2.1 eV and a ferroelectric phase fraction 10% lower than that of the 19H devices, yielding an RMSE of 1.6, shown in **Figure 3a**. The 1.5 eV defect energy fits well with the expected F-D defect energy values and lies within the range of defect energies expected from photoluminescence and cathodoluminescence measurements (**Figure S6**).[38]

To illustrate the impact of tunneling on ferroelectric switching, **Figure 3b** compares the simulated switching current density of the simulated interlayer device with and without electron tunneling. Electron transport across the $Al_2O_3$ layer, modeled as tunneling, is necessary to produce the enhanced internal electric field in the interlayer device structure. The contribution of the mobile charges creates an internal field, which is marked in the current-electric field plots given in **Figure 3b** as $E_{int}$. Electron transfer across the interface creates a bidirectional internal electric field that depends on the electrical history of the device, consistent with the theory of fluidic imprint.

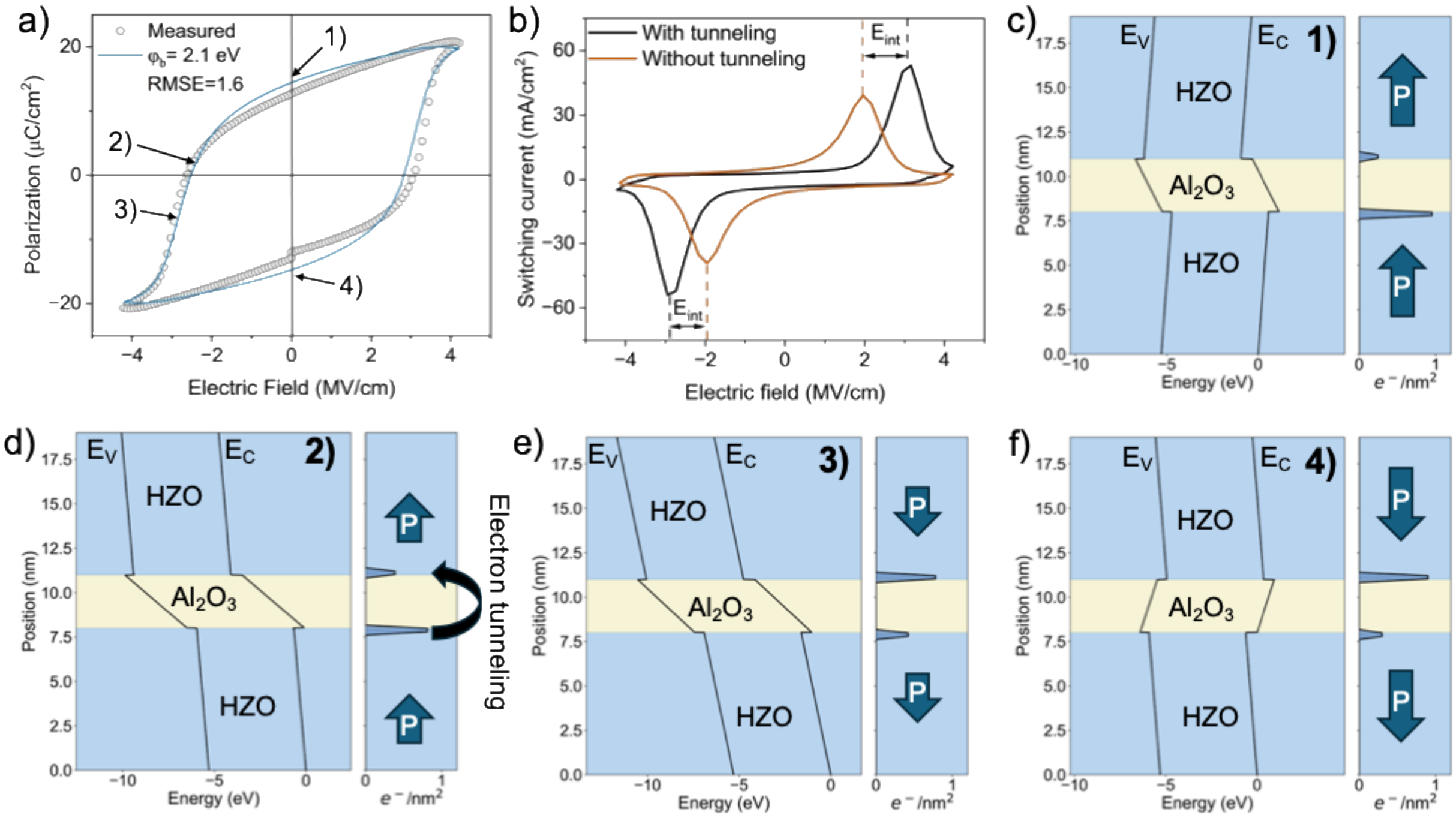

**Figure 3. Modeling of defect-mediated coercive field modulation in interlayer ferroelectric devices and experimental validation. 3a**. Phase-field simulation of ferroelectric switching in an 8H-3A-8H device under fluidic imprint conditions compared to experimental data; **3b**. Comparison of the switching current of the 8H-3A-8H device with and without tunneling; **3c-f**. Band structures measured from XPS/REELS are used to construct schematic band diagrams and compared to the phase-field simulation data of the 8H-3A-8H stack at points 1-4 in the ferroelectric hysteresis, marked in (**3a**). The gaussian charge

distributions to the right of each band diagram illustrate the calculated interfacial electron accumulation.

To visualize how ferroelectric hysteresis modifies internal fields and defect distributions, band diagrams of the interlayer device at different points in the switching cycle are shown in **Figure 3c-f.** These diagrams were constructed using the experimentally determined band alignment and the electrostatic potential from the phase-field simulations. The calculated density of electrons at each F-D interface is represented by the peaks in the figure to the right of each band diagram. The band diagram for the $+P_r$ state (marked as point 1) is shown in **Figure 3c**. In this $+P_r$ state, the electrons accumulate at the bottom F-D interface, producing a positive internal field across the ferroelectric and a negative field across the $Al_2O_3$. The calculated electric fields are 0.56 MV/cm and -3.72 MV/cm for the HZO and $Al_2O_3$, respectively. As the external applied electric field approaches $-E_c$, the polarization begins to switch and the electrons tunnel from the bottom interface to the top interface, as demonstrated in **Figure 3d** for point 2. As the electric field is driven more negative to point 3, the polarization is reoriented and electrons have tunneled across the dielectric layer, displayed in **Figure 3e**. Finally, at point 4 the material is left in a $-P_r$ state after removing the applied electric field, portrayed in **Figure 3f**. At point 4, the electrons have redistributed to the top F-D interface and reversed the internal field direction compared to $+P_r$. The evolution of electron density and internal fields in the 8H-3A-8H device architecture demonstrates that the phase-field model accounts for the electron transfer across the dielectric, as well as the electrostatic coupling from the change in interfacial electron accumulation. These band diagrams and electron accumulation derived from the simulated device show that: **1)** The internal electric field reverses with polarization switching due to the redistribution of electrons **2)** electrons accumulate at the top F-D interface with a $-P_r$ and at the bottom F-D interface with a $+P_r$. The variable internal field that occurs only when defects are added to the model shows a clear connection between defects and fluidic imprint in HZO interlayer devices.

**Figure 4a** provides a graphical summary from the calculated data of the change in simulated interlayer charge density and polarization as a function of a time-variable applied external electric field. Under an applied electric field, electrons redistribute across the $Al_2O_3$ layer, producing internal bias fields ($E_{int}$) through charge accumulation at the F–D interfaces. The subset of **Figure 4a** shows how the calculated $E_{int}$ varies as a function of time under an external field. The electric field in the ferroelectric ($E_{FE}$) is non-zero when the applied field is zero, noted by $E_{Int}$.

Experimentally measured first-order reversal curve (FORC) measurements provide further corroboration on the presence and impact of a variable internal field in the interlayer devices. The waveform used in FORC measurements to extract the switching density is shown in **Figure S10a,** and an example of the J(E) hysteresis taken on a control device is shown in **Figure S10b**. The P(E) hysteresis for the control and interlayer devices from FORC measurements are shown in **Figure S10c** and **Figure S10d**, respectively. The FORC switching density plots for 19H and 8H-3A-8H devices are shown in **Figure 4b** and **Figure 4c**, respectively. Both the applied electric field and reversal fields at which the switching density is highest are larger in the 8H-3A-8H samples compared to the 19H samples, consistent with the higher coercive field of the 8H-3A-8H samples. Using the electric field and reversal field with the largest switching density marked on each FORC distribution, the internal field ($E_{int}$) can be calculated by:

$$E_{int} = \frac{E + E_R}{2} \quad \textbf{Eq. 1}$$

where E and $E_R$ are the electric field and reversal field at the maximum switching density, respectively.[52] The FORC density distributions are compared to the FORC distributions simulated using the defect-driven phase-field model. The modeled results are overlaid on the experimental data in **Figure 4b**, and **4c** for the 19H and 8H-3A-8H, respectively. The simulated FORC density distributions are marked on the plots with dashed contours representing switching density ≥80% of the maxima.

Remarkably, the difference in $E_{int}$ between the control and interlayer samples (-0.56 MV/cm) closely matches the calculated internal field across the ferroelectric layer from the phase-field simulations (-0.55 MV/cm). The calculated bias field for 19H is $E_{int}$ = -0.23 MV/cm, while that for 8H-3A-8H is significantly larger at $E_{int}$ = -0.79 MV/cm. The simulated FORC distribution of the 19H overlaps the majority of the experimental data. The point of maximum switching density is within 6% for both electric field and reversal field. The difference in $E_{int}$ differs by 0.01 MV/cm. For the interlayer devices, the reversal field is within 5% of the experimental reversal field, while the electric field shows a larger deviation (15.3%). The deviation between the measured and simulated FORC for the interlayer sample is attributed to the presence of other charge transport mechanisms (e.g., trap-assisted tunneling, Poole-Frenkel conduction) that become more significant under the reduced fields used for FORC.[53] Overall, the FORC data corroborate the hypothesis that interfacial defects mediate the increase in coercive field.

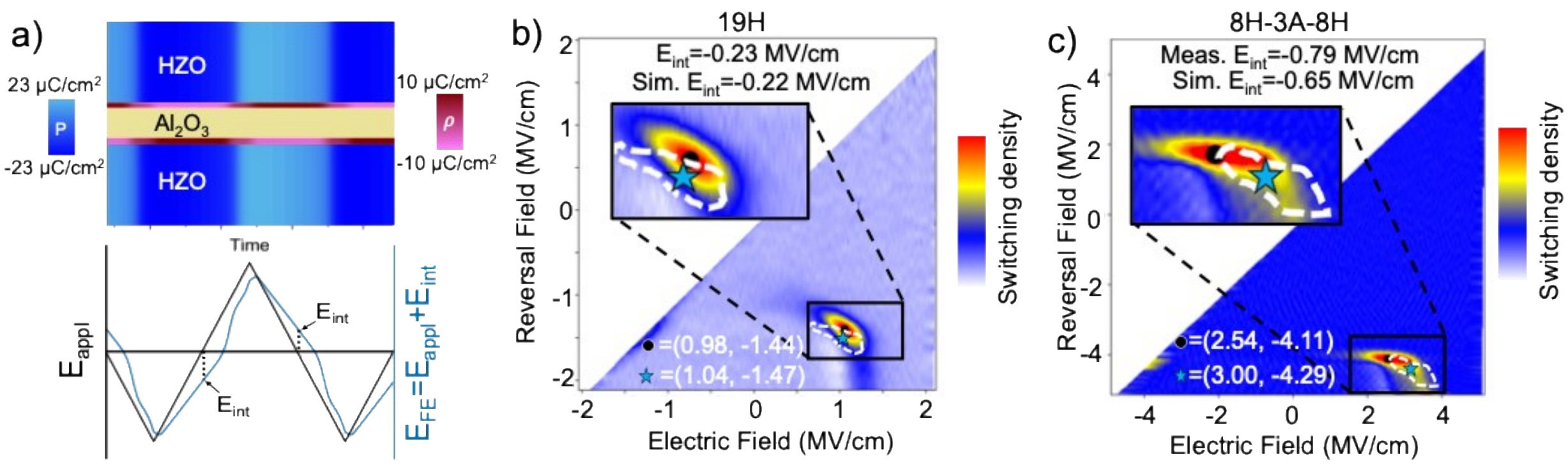

**Figure 4. Fluidic imprint and internal bias fields in interlayer HZO devices. 4a**. Graphical summary of the simulated time dependent polarization (P) and interfacial charge density (ρ) in the device as the applied electric field ($E_{Appl}$) in the 8H-3A-8H model. The electric field in the ferroelectric ($E_{FE}$) is non-zero when the applied field is zero, noted by $E_{int}$; **4b.** The modeled and measured FORC distribution of control 19H HZO; **4c.** The modeled and measured FORC distribution for the 8H-3A-8H interlayer HZO, with both experimental (heat map) and simulated (dashed lines) results. The points of maximum switching density are labeled with a circle and star for experimental and simulated data, respectively.

The agreement between the simulated and measured P(E) hysteresis and the internal fields and switching distributions determined from FORC provides indirect but compelling evidence that defect-mediated electrostatics govern the observed modulation of ferroelectric switching. To further evaluate this hypothesis, direct evidence of interfacial defect states and charge redistribution is needed. A key finding from the defect-dependent phase-field model is that the occupancy of defect states changes with poling. The internal fields calculated in the phase-field simulations and measured by FORC provide an opportunity to directly probe the interlayer fluidic imprint, as internal fields in layered thin films can be directly measured via XPS core-level shifts.[54,55] A positive internal electric field increases the measured binding energy, as the electrical potential reduces the kinetic energy of emitted electrons. Based on the modeled band diagrams, the Al 2p peak in the -Pr state should therefore shift to higher binding energy relative to the +$P_r$ state. To test this prediction, interlayer samples were electrically cycled and poled into a +/- polarization state using removable top electrodes. Examples of the P(E) loops taken throughout the cycling and the conditions used for poling are presented in **Figure S11**. A schematic of the poling process is shown in **Figure 5a**, in which removable indium contacts are used to pole the devices in either a +$P_r$ or -$P_r$ state.  The polarization and bias directions are referenced from the bottom electrode, consistent with **Figure 3a**. The top electrode was then mechanically removed, and the poled regions were probed via XPS after etching to the interface.

No significant changes in the Hf 4f, Zr 3d, or O 1s spectra were observed with poling, shown in **Figure S12a-c**, indicating that the oxygen vacancy concentration in the bulk remains stable with polarization switching.[34,43] In contrast, the peak position of the Al 2p peak, shown in **Figure 5b**, shifts by approximately 250 meV to higher binding energy when poled in the -$P_r$ state relative to the +$P_r$ and as-grown states. **Figures S14a-c** demonstrate that the other metal oxide peaks are stable within ±50 meV while the aluminum peak shifts by approximately 250 meV compared to the +$P_r$ and unpoled samples. The internal field across the $Al_2O_3$ is ~7 times larger than that across the HZO layers, which is why the Al 2p peak exhibits a larger shift. To exclude the possibility of a rigid shift caused by polarization switching,[56] core-level peaks are also analyzed on poled samples before etching and charge-shifted to the adventitious carbon (C 1s) peak at 284.8 eV, shown in **Figure S14d-f**. In all cases, the difference in peak binding energy is within ±50 meV, confirming that the shift in the Al 2p peak is caused by an internal electric field rather than a rigid shift due to polarization. The shift in the Al 2p peak provides further evidence for the presence of a polarization-dependent internal electrical field across the dielectric layer, consistent with the predictions from the defect-based phase-field model and the proposed fluidic imprint mechanism.

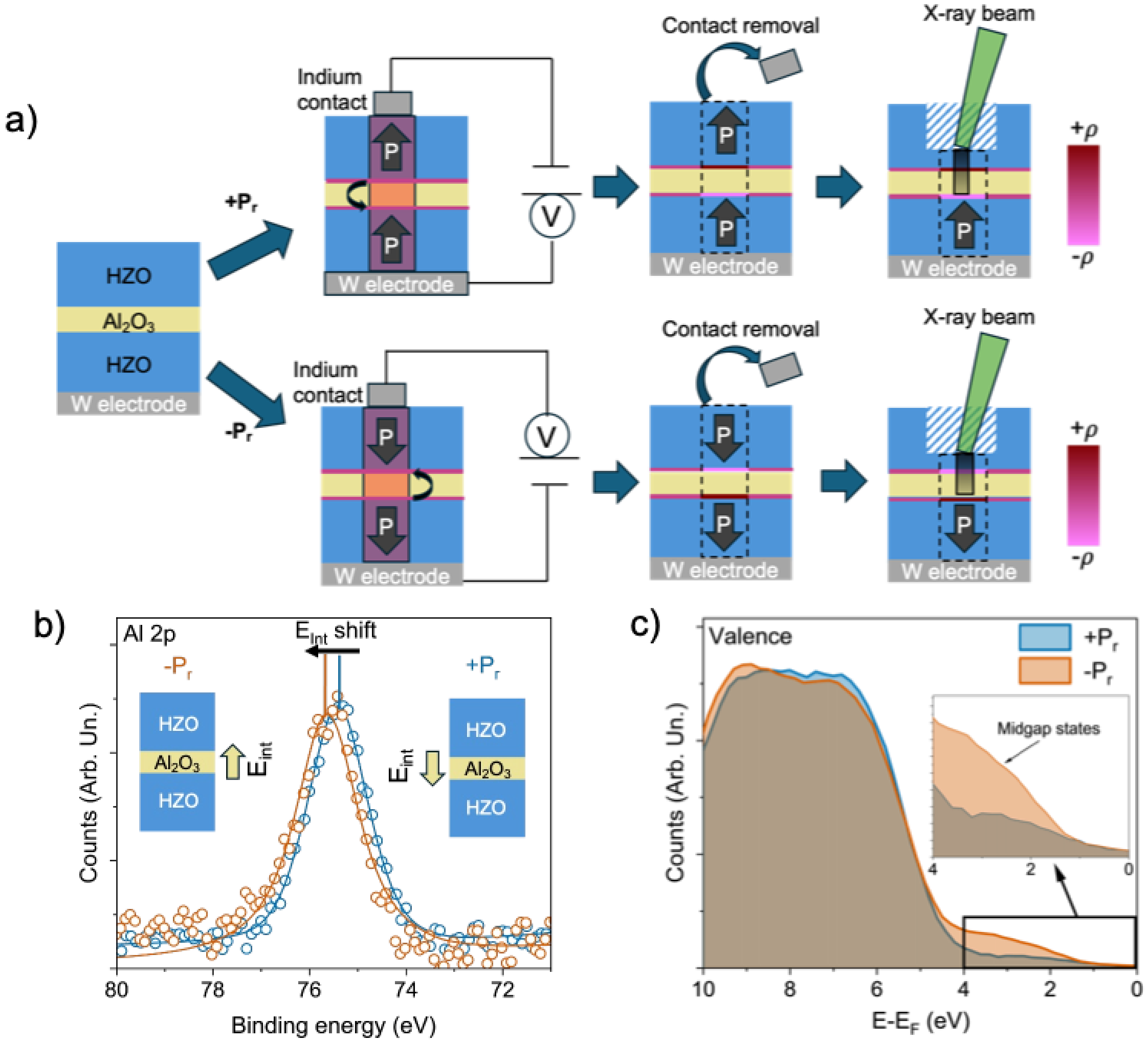

**Figure 5. Simulation and experimental evidence of polarization-dependent internal fields and charge transport in 8H-3A-8H device stacks. 5a.** Schematic of the procedure used to pole the interlayer devices with a positive(+$P_r$) or a negative (-$P_r$) remanent polarization and subsequently measure the poled regions via XPS; **5b.** Al 2p core-level XPS after etching to the $Al_2O_3$ interface for samples poled into +$P_r$ and -$P_r$ states; **5c.** Valence band spectra of the same regions shown in **(b).**

While core-level spectra show no measurable change in oxygen vacancy concentration with poling (**Figure S12a-c**), valence band spectra are sensitive to the electronic occupancy of defect states.[43,57,58] The phase-field model predicts an increased electron accumulation at the top F-D interface in the -$P_r$ state, which would appear in the valence spectra as it is dominated by the top interface, with a ~3:1 ratio determined by the Beer-Lambert law.[59] Therefore, the predicted increase in electron accumulation would reveal itself as an increased spectral weight of midgap states.[43,57,58] **Figure 5c** shows the polarization-dependent valence spectra for a representative interlayer device. The increase in midgap peak intensity in the $-P_r$ state compared to the $+P_r$ and as-grown states is consistent with the predicted change in electron density from the phase-field model. Both the core-level and valence spectra closely match the predicted charge accumulation across the $Al_2O_3$ layer.

To assess the cycling stability of the defect states, an additional region was cycled for either 10 or 100 times and left in the $+P_r$ state. The Al 2p core-level and valence spectra of the as-grown, 10-cycle $+P_r$, and 100-cycle $+P_r$ regions are nearly identical (**Figure S15a,b**), indicating that cycling does not significantly alter the defect distribution or electronic structure at the interface. While oxygen vacancy migration has been shown to degrade lifetime,[60] the oxygen vacancies here are spatially fixed at the interface and therefore reshape the electrostatic boundary condition through electronic compensation without inducing long-term instability. Kelvin probe force microscopy (KPFM) probed the temporal stability of this electronic compensation (**Figure S13**). Subcoercive tip-bias patterning at opposite polarities produced regions of surface-potential contrast following the sign of the applied bias, consistent with charge injection and trapping. The surface potential of each region was monitored over several hours as it relaxed toward the unwritten state value, and the relaxation fitted to extract a decay time constant. The 8H-3A-8H sample exhibits slower charge relaxation than the 19H sample for both polarities ($\tau$ = 62.2 vs 49 min for the +1 V region and 167 vs 83.4 min for the -1 V region). Slower relaxation is consistent with deeper or more stable trapping states, reduced charge-transport pathways, or slower defect-mediated charge redistribution, and suggest that screening of the depolarization field may be sustained longer with the insertion of the dielectric layer.

Collectively, these results further support that the dominant response is due to polarization-dependent electronic filling of defect states rather than oxygen vacancy migration, which is consistent with prior research that has shown that dielectric interlayers improve device lifetime.[11] Overall, the XPS core-level and valence measurements reveal three key features of the interlayer devices: 1) There is an increased concentration of charged defects (oxygen vacancies) at the F-D interface; 2) The internal field across the dielectric reverses with polarization direction; 3) The switching-dependent internal field arises from electronic redistribution of defect states rather than migration of oxygen vacancies.

### 2.5 Mechanisms of polarization switching

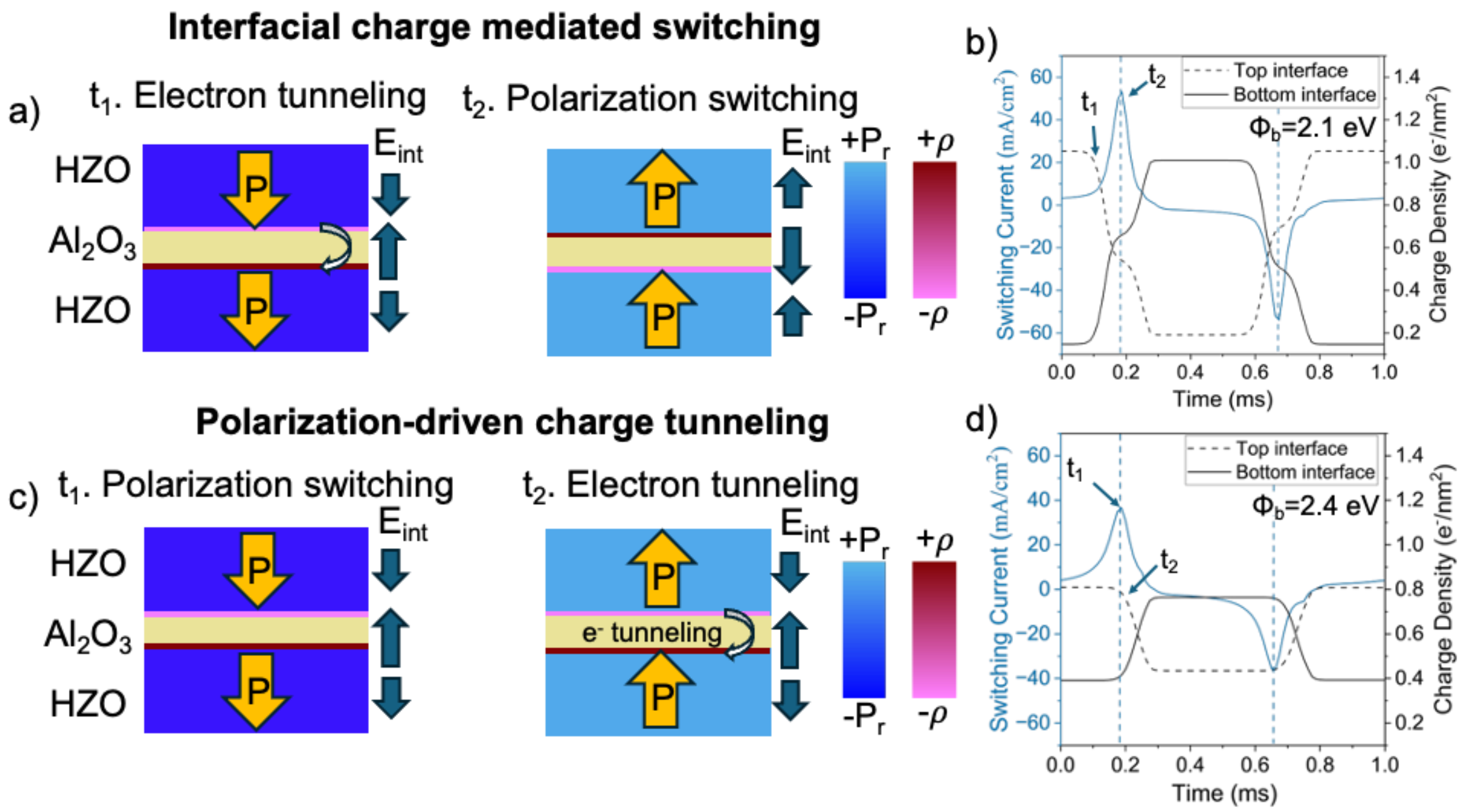

**Figure 6. Defect-mediated switching mechanisms in interlayer HZO devices. 6a** Graphical depictions of the time dependence ($t_1$, $t_2$) of the interfacial charge-mediated switching mechanism in HZO; **6b**. Phase-field simulation data showing interfacial charge-mediated switching occurs with a barrier of 2.1 eV**; 6c.** Depiction of the time dependence of the polarization-driven charge tunneling switching mechanism in interlayer HZO devices; **6d**. Phase-field simulation data of polarization-driven charge tunneling with a tunnel barrier of 2.4 eV.

Building upon the defect-mediated ferroelectric switching model, we identify two distinct pathways through which defects modulate polarization switching:

**Type 1. Interfacial charge-mediated switching:** At sufficiently high electric field, electrons tunnel across the dielectric barrier, changing the direction of the internal field, assisting in the polarization reversal (**Figure 6a**).

**Type 2. Polarization-driven charge tunneling:** The applied electric field reaches $E_{appl}>E_c+E_{int}$, at which point the polarization direction will switch, changing the electrostatic landscape of the device and therefore promoting the transfer of electrons after switching (**Figure 6c**).

Understanding the impact each switching pathway has on the ferroelectric properties of interlayer devices is critical. Phase-field simulations show that the dominant switching mechanism depends on the tunneling barrier. For $\varphi_b$=2.1 eV, the simulated charge redistribution precedes polarization reversal, shown in **Figure 6b**, indicating a defect-driven mechanism. Increasing the barrier to 2.4 eV, as in **Figure 6d**, suppresses early electron transfer, shifting the device into a polarization-driven charge tunneling regime. The maximum tunneling current ranges from $10^{-4}$-$10^{-3}$ A/cm$^2$, which is slightly lower than previously measured tunnel current density through 3 nm $Al_2O_3$ for M-I-M and M-I-S devices.[61,62]

In order to translate our findings into actionable directives for device design we connect the defect-modulated switching models to the directly controllable variables: the interfacial defect density, and dielectric thickness. Simulations of 36 devices with varied charge densities and dielectric thickness with a constant device thickness of 19 nm are shown in **Figure 7**. A clear boundary emerges based on the coercive field and switching pathway. The highest predicted coercive field for an interlayer HZO device is for a 7.5H-4A-7.5H device with an interfacial charge density of 0.6 e$^-$/nm$^2$, with a simulated coercive field of 3.4 MV/cm. The phase-field simulations reveal that the maximum coercive field, and memory window, occurs when the switching is at the barrier between charge-mediated and polarization-driven switching mechanisms. These critical insights demonstrate how to tune the interface-driven modulation of the ferroelectric properties to the design of HZO for FeNAND. Additionally, these models can be readily translated to a wide range of ferroelectric-dielectric systems.[1] Phase-field modeling shows critical considerations for device design are the band alignment, the propensity for interfacial charged defects, and the thickness of dielectric material. These results exemplify that understanding the mechanism of ferroelectric switching is critical for designing the next-generation ferroelectric-interlayer memory devices.

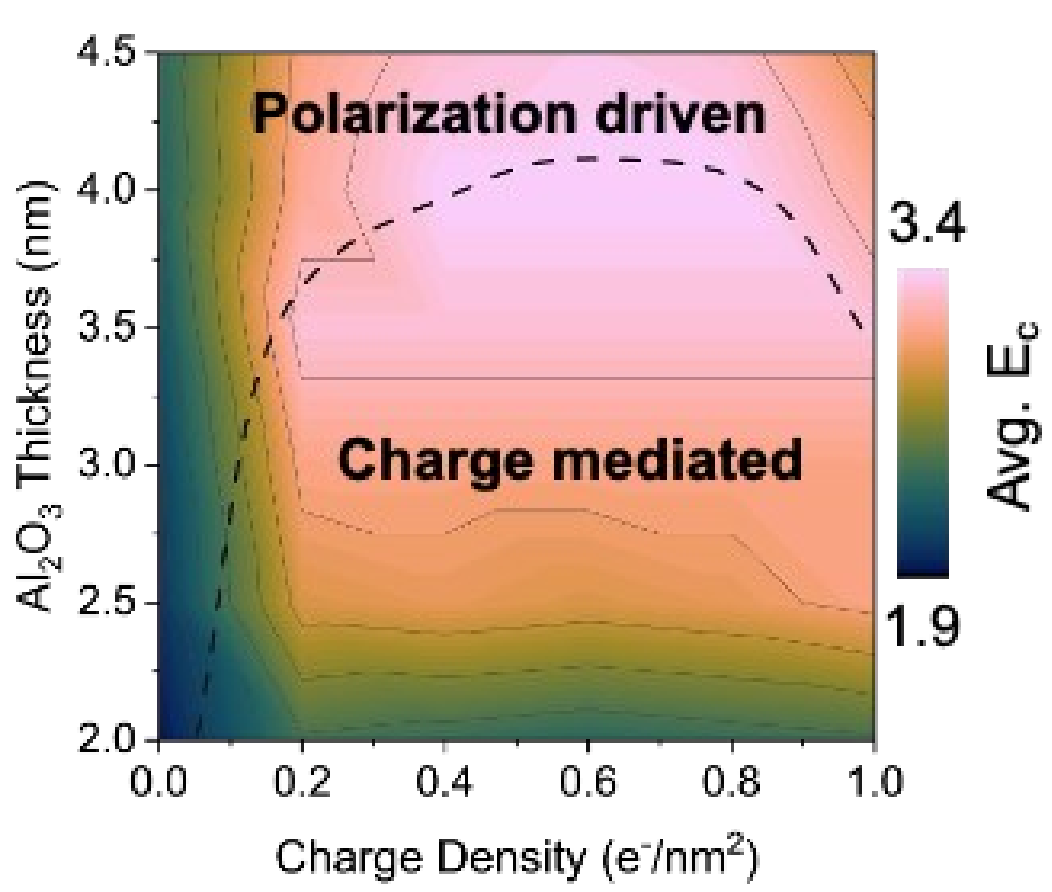

**Figure 7.** Simulated coercive field of devices as a function of interlayer thickness and barrier height. The dashed line separates regions of interfacial charge-mediated switching and field-driven intrinsic switching.

## 3. Conclusion

Inserting a dielectric interlayer produces defects that create a reversible internal electric field that impacts the polarization switching and increases the coercive field by more than threefold. Polarization-dependent XPS measurements combined with first-order reversal curve analysis reveal that switching is governed by electronic compensation of stationary defects rather than defect migration. Building from the measured P(E) hysteresis and band structure, a defect-dependent phase-field model is developed, which shows the defect-mediated switching pathway for HZO. The model developed in this work provides a predictive framework for electrostatic control of ferroelectric switching through interfacial defects and supports interface-controlled switching as a scalable strategy for high-density 3D ferroelectric memory. These findings redefine the role of defects in ferroelectric HZO from deleterious to engineerable, enabling direct control of ferroelectric switching.

## 4. Acknowledgements

This research was supported by the Center for 3D Ferroelectric Microelectronics Manufacturing (3DFeM$^2$), an Energy Frontier Research Center funded by the U.S. Department of Energy, Office of Science, Basic Energy Sciences under Award No. DE-SC0021118. Scanning electron microscope cathodoluminescence was supported by the Center for Nanophase Materials Sciences (CNMS), which is a US Department of Energy, Office of Science User Facility at Oak Ridge National Laboratory. Lance Fernandes' work on the fabrication of HZO/$Al_2O_3$ heterostructures was supported by Samsung Electronics (Project No. IO250304-12193-01). This work was performed in part at the Georgia Tech Institute for Electronics and Nanotechnology (IEN). The IEN is a member of the National Nanotechnology Coordinated Infrastructure (NNCI), which is supported by the National Science Foundation (Grant ECCS-1542174). The authors would also like to acknowledge Muhammad Islam, Tanay Patni, and Prof. Azad Naeemi for fruitful discussions concerning phase-field modeling of HZO.

## Methods

*Thin Film Synthesis and Device Nanofabrication*

$Hf_{0.5}Zr_{0.5}O_2$ films were deposited by atomic layer deposition (ALD) using a Veeco Fiji ALD tool. The substrates of 19H and 8H-3A-8H samples are degenerately doped silicon with phosphorus (0.005 Ohm-cm). The substrates of other interlayer samples with varying $Al_2O_3$ thickness are degenerately doped silicon with boron (0.005 Ohm-cm). The bottom tungsten electrode is deposited by sputtering at room temperature with 50 mTorr argon atmosphere. The HZO films were deposited using the hafnium precursor tetrakis(dimethylamido)hafnium (TDMaHf) and the zirconium precursor is tetrakis(dimethylamido)zirconium (TDMaZr). The $Al_2O_3$ dielectric interlayers were also deposited by ALD using trimethyl aluminum (TMA) precursor. The deposition temperature for both the $Al_2O_3$ and $Hf_{0.5}Zr_{0.5}O_2$ layers is held at 250 °C. The precursor pulse time is 0.25 and 0.03 seconds for Hf/Zr and Al precursor respectively. The carrier and purge gas used in the ALD chamber is argon. The purge time is 15 seconds. The thickness of the layers is controlled by the number of ALD cycles at 53 and 38 for the 8 nm HZO and 3 nm $Al_2O_3$ respectively. The number of cycles for 1 nm, 5 nm, and 10 nm $Al_2O_3$ was 15, 70, and 130 cycles, respectively. The top electrodes were deposited via the same sputtering process used for bottom electrode. After top tungsten deposition, the films were annealed at 500 °C for 30 seconds in nitrogen atmosphere. Tungsten was selected as the top and bottom electrode material because it increases the remanent polarization of HZO compared to other electrode materials due to its high elastic modulus.[63]

*Structural characterization*

The crystal structure of the films was characterized by X-ray diffraction (XRD) and scanning transmission electron microscopy (STEM). The XRD was collected using a Rigaku Smartlab XE using a parallel beam configuration. The 2θ measurements were collected using a grazing incident geometry, with $\alpha$=0.75°. The measurement range was 25° to 34°, measured at 0.5°/min. Scanning transmission electron microscopy (STEM) images were collected with an aberration corrected STEM Hitachi HD2700 using an annular dark field detector working at 200 kV with a spatial resolution of 0.13 nm. STEM samples were prepared using a Thermo Fisher Helios 5CX focused ion beam (FIB). The final polishing condition was performed at 43 pA and 5 kV.

*Electrical characterization*

Polarization-electric field (P(E)) hysteresis, and current density-electric field (J(E)) hysteresis, were measured with a Radiant Precision Premier II ferroelectric tester. Unless otherwise noted, the P(E) hysteresis measurements were taken at a frequency of 100 Hz. The averages and errors for the coercive fields of each device were calculated using the standard error from ten electrodes per device type.

First-order reversal curves (FORC) were measured with a Radiant Precision Premier II ferroelectric tester. FORC is an established method for determining internal bias fields in HZO and other ferroelectric materials.[64–66] FORC consists of a series of P(E) hysteresis measurements with electric fields cycled between a fixed positive saturation field and a variable reversal field.[67,68] FORC was measured from 4.5 MV/cm to -4.5 MV/cm for the 8H-3A-8H devices and from 2.1 MV/cm to -2.1 MV/cm for the 19H devices using a reversal field step size of ~0.1 MV/cm. The time per step for the FORC measurements was 3 ms. The FORC density graphs were calculated and plotted using a custom Python code using the Spyder (Version 6) platform. Using the FORC data, the Preisach switching density for ferroelectric materials is defined as:

$$\rho^{-}(E_r,E)=\frac{1}{2}\cdot\frac{\partial^2 P^{-}(E_r,E)}{\partial E_r \partial E}=\frac{1}{2\dot{E}}\cdot\frac{\partial j^{-}(E_r,E)}{\partial E_r} \qquad \textbf{Eq.2}$$

where $\rho^{-}(E_r,E)$ is the switching density for the low-to-high sweep at a given reversal field ($E_r$) and applied field (E), P is the polarization, and j is the current density. The switching density is approximated as:

$$\rho^{-}(E_r,E)\approx\frac{1}{2\dot{E}}\cdot\frac{j^{-}(E_{r,i},E)-j^{-}(E_{r,i-1},E)}{E_{r,i}-E_{r,i-1}} \qquad \textbf{Eq. 3}$$

The switching density was extracted using a custom code using the Spyder (Version 6) platform.

The capacitance-voltage hysteresis response was measured with a Modulab XM MTS system (Solartron) with a femtoammeter module. A triangular waveform was used, with an amplitude of 3.68 MV/cm for 8H-3A-8H and 1.58 MV/cm for 19H and a period of 240 s. The AC amplitude for CV measurements was 100 mV.

*Chemical Characterization*

X-ray photoelectron spectroscopy (XPS) measurements were taken with a NEXSA G2 spectrometer with an aluminum K-$\alpha$ source (1.486 keV), a flood gun, and 180°, double-focusing hemispherical analyzer with a 128-channel detector. The XPS spectra were fit with Avantage software (Version 6.7.0). The Lorentzian/Gaussian mix was constrained below 65% and the FWHM was held below 2 eV. The oxygen peaks fit with a metal oxide peak at 530.2 eV and a non-lattice oxygen peak at 531.7 eV, which is likely due to adsorbed oxygen.[43] Both the zirconium and the hafnium peaks are fit with 4+ and 3+ oxidation state doublets using a procedure described in reference [69], with both suboxide peaks separated by 1-1.2 eV from the 4+ peaks. The aluminum peak is fit with a single peak at 74.2 eV, which is within the range of measured values reported for ALD $Al_2O_3$.[70] The valence spectra were collected with an electrical sample holder with a -10 V bias between the sample holder and the detector to measure the cutoff electrons and mid-gap states, as is standard in these measurements.[71] The bottom electrode was grounded to the sample holder using a metallic pin. The Fermi energy of the sample holder and thin film are expected to be equal in these measurements.

The effect of poling on chemical states at the $Al_2O_3$-$Hf_{0.5}Zr_{0.5}O_2$ interface was measured by poling samples with a removable indium contact. The contacted area was marked to define the poled area, and indium contacts were removed. Samples were cycled using a typical double bipolar triangular waveform with a 4.74 MV/cm amplitude and a frequency of 100 Hz. The effective electrode area was calculated using the known relative permittivity and polarization of samples with defined electrode areas and the measured capacitance and polarization of the contacted area. In each case, the effective electrode area was greater than 0.16 $mm^2$, larger than the area measured by XPS. All peaks after etching are charge-shift corrected using the reference for the W 4f surface peak (**Figure S16a-c**) to ensure that charging did not cause an artificial peak shift. A comparable peak shift was observed when charge-shift referencing to the primary O 1s peak.

Due to the limited measurement depth of XPS (~2-3*attenuation length), etching is necessary to measure the $Hf_{0.5}Zr_{0.5}O_2$-$Al_2O_3$ interface. Etching was performed with an argon gas cluster source, with a cluster size of 2,000 atoms and a cluster energy of 8,000 eV. To measure the change in composition at the HZO-$Al_2O_3$ interface by XPS, an ion cluster etch was used to suppress ion implantation and reduction of the metal-oxides compared to a monatomic Ar ion etch according to reference [72]. The X-ray spot size was 0.2 mm and the ion beam spot size was 2 mm. The time needed to etch to the top F-D interface of the 8H-3A-8H sample is 1,600 seconds, shown in **Figure S5**. In 8H-3A-8H, the Al 2p peak is measured above the noise level after 520 seconds of etching, with further etching increasing the measured composition of aluminum linearly. After approximately 1,600 seconds of etching, the measured composition of the aluminum is comparable to the other cations. Ion etching is also conducted on the 19H devices under the same conditions to provide a baseline for the amount of defect imparted by etching for comparison. The thickness of the HZO overlayer after etching was calculated using:

$$\frac{I_A}{I_B} = R = R_\infty \frac{\left[1 - \exp\left(-\frac{d}{(\lambda_{\{A,A\}}\cos\theta)}\right)\right]}{\exp\left(-\frac{d}{(\lambda_{\{B,A\}}\cos\theta)}\right)} \quad \textbf{Eq. 4}$$

where $I_A$ and $I_B$ are the intensities of the overlayer (HZO) and underlayer ($Al_2O_3$), respectively, $R_\infty$=0.225 (0.173) for Hf 4f (Zr 3d) peaks, $\theta$ =0°, and d is the overlayer thickness. The attenuation length in each material was calculated according to the Tanuma-Powell-Penn-2M method,[73] yielding effective attenuation lengths of $\lambda_{\{A,A\}}$=1.736 nm, $\lambda_{\{B,A\}}$=1.678 nm, and $\lambda_{\{B,B\}}$=2.68 nm. The thickness of the HZO overlayer is determined to be 3.55 nm and 3.42 nm after etching using the Hf 4f and Zr 3d core-level peaks, respectively. Next, the relative intensity of the valence spectra from the top and bottom interfaces was determined using the Beer-Lambert law, yielding a 2.84:1 top interface:bottom interface intensity ratio.[59]

The defect state and band structure were further corroborated using photoluminescence (PL). PL measurements were performed using an Invia Qontor Renishaw dual Raman/photoluminescence system. The excitation wavelength was 488 nm (2.5 eV) with a grating of 2400 lines/mm. The photoluminescence data was averaged across three randomly selected regions on each sample type. Cathodoluminescence (CL) measurements were performed at both room temperature and 10 K using an FEI Quattro environmental SEM equipped with a Delmic Sparc CL collection module. A parabolic mirror was used to collect CL signals emitted from the film under electron-beam excitation, with the beam directed through an aperture in the mirror onto the sample surface. Spectra were acquired with a 2-second integration time per pixel across a raster scan spanning several micrometers, then averaged to produce statistically representative CL spectra. Newton BEX2-DD detector is used, which has a spectral range cutoff below 350 nm.

*Band diagram measurements*

The band gap, valence band maxima, Fermi energy, and work function of HZO and $Al_2O_3$ were directly measured in order to accurately model the tunneling mechanics through the interlayer. The ionization potential ($E_f$-$E_V$) of a material was measured using the valence band spectra from XPS, shown in **Figure S8a,e**. Furthermore, the work function ($E_{Vac}$-$E_F$) can be measured by analyzing the cutoff spectra, shown in **Figure S8b,f**. The cutoff spectra were measured in-situ with the valence band spectra and are also measured with a -10 V bias applied between the stage and detector to measure the low kinetic energy electrons. Finally, the excitonic band gap of a material was measured as the difference between the energy of elastically and inelastically scattered electrons via reflection electron energy loss spectroscopy in **Figure S8c,g**. From these three measurements, the band structure of the 8H-3A-8H stack was recreated, which was used to determine the tunneling barrier heights for the phase-field models.

*Phase-Field Modeling*

Phase-field modeling was performed to determine the effect of defects on the ferroelectric switching of $Hf_{0.5}Zr_{0.5}O_2$. The phase-field modeling was performed through the FERRET module[31] based on the Multiphysics Object Oriented Simulation Environment.[32,33] Parts of the phase-field modeling code were modified from a large language model (Claude Opus 4.6, Anthropic), and were subsequently reviewed, modified, and validated by the authors. Phase-field models were simulated using a 60 nm x 19 nm system size with a mesh size of 0.6 nm in the x-direction and 0.19 nm in the y-direction. The 19H architecture was modeled as a single 19 nm $Hf_{0.5}Zr_{0.5}O_2$ (HZO) layer. The parameters for the time-dependent LGD model are determined using the measured 19H device P(E) hysteresis and capacitance-voltage hysteresis loop (given in **Figure S17a,b**). Both the top and bottom HZO layers of the 8H-3A-8H device architecture were modeled using 7.5 nm of ferroelectric HZO, 0.5 nm ferroelectric HZO as a charge trap layer at the interface of the $Al_2O_3$ and a 3 nm interlayer of $Al_2O_3$. The HZO switching parameters were the same as the experimentally derived characteristics of the 19H control device. The ferroelectric switching was modeled using the time-dependent Landau-Ginzburg-Devonshire equation,[74]

$$\frac{\partial P}{\partial t} = -\Gamma_P \frac{\delta F}{\delta P} \qquad \textbf{Eq. 5}$$

where P is the polarization of the system, $\Gamma_P$ is the kinetic coefficient (2.9 x $10^{-6}$ $\Omega^{-1}cm^{-1}$), and F is the system Helmholtz free energy. The kinetic coefficient was calibrated to match the experimental P(E) hysteresis loops. In the developed phase-field model, the free energy of the system is composed of,

$$\mathrm{F} = \iiint_V \mathrm{f}_{bulk} + f_{\nabla P} + f_{elec} + f_{elastic} d^3 r \qquad \textbf{Eq. 6}$$

where $f_{bulk}$ is the double-well potential energy, $f_{\nabla P}$ is the gradient energy density for the formation of domain walls, $f_{elec}$ is the interaction energy of an internal or externally applied electric field, and $f_{elastic}$ is the coupling between the dipole moment and strain. The bulk free energy ($f_{bulk}$) was defined as,

$$f_{bulk} = \alpha_1 {P_1}^2 + \alpha_{11} {P_1}^4 + \alpha_{111} {P_1}^6 \qquad \textbf{Eq. 7}$$

where $P_1$ is the out-of-plane polarization and $\alpha_1$, $\alpha_{11}$, and $\alpha_{111}$ are the Landau coefficients for HZO. Because there is a wide range of reported Landau parameters for HZO, the 19H experimental results were estimated using a custom code through Spyder (Version 6). The experimentally determined results are $\alpha_1$= -2.16 x $10^8$ m/F, $\alpha_{11}$= 6.92 x $10^9$ $m^5/FC^2$, and $\alpha_{111}$ = 9.29 x $10^9$ $m^9/FC^4$. The gradient energy density was calculated via:[75]

$$f_{\nabla P} = \frac{G_{11}}{2}\left(P_{x,x}^2 + P_{y,y}^2 + P_{z,z}^2\right) + G_{12}\left(P_{x,x}P_{y,y} + P_{y,y}P_{z,z} + P_{x,x}P_{z,z}\right) + \frac{G_{44}}{2}\left[(P_{x,y}P_{y,x})^2 + \left(P_{y,z}P_{z,y}\right)^2 + \left(P_{x,z}P_{z,x}\right)^2\right] \quad \textbf{Eq. 8}$$

where $G_{ijkl}$ is the gradient energy coefficient and $P_{i,j}$ is the partial derivative $\frac{\partial P_i}{\partial x_j}$. The gradient energy coefficient $G_{110}$ was set as 5.066 x $10^{-10}$ $Jm^3/C^2$ and the relative gradient energy coefficients were set as $G_{11}/G_{110}$=1, $G_{12}/G_{110}$=0, and $G_{44}/G_{110}$=0.5 from reference [16]. The electrostatic interaction between polarization and the electric field is given by:

$$f_{elec} = P_j\frac{\partial V}{\partial x_j} - \frac{1}{2}\varepsilon_0\varepsilon_r E_i E_i \quad \textbf{Eq. 9}$$

where V is the electrostatic potential, $\varepsilon_0$ is the permittivity of free space (8.85 x $10^{-12}$ F/m) and $\varepsilon_r$ is the relative permittivity. The electrostrictive energy density accounts for the mechanical strain, given by:

$$f_{elastic} = \frac{1}{2}C_{ijkl}\left(\varepsilon_{ij} - \varepsilon_{ij}^0\right)\left(\varepsilon_{kl} - \varepsilon_{kl}^0\right) \quad \textbf{Eq. 10}$$

where $C_{ijkl}$ is the elastic stiffness tensor, $\varepsilon_{ij}$ and $\varepsilon_{kl}$ are the total strain tensor components, $\varepsilon_{ij}^0$ and $\varepsilon_{kl}^0$ are the electrostrictive eigenstrains, given by:

$$\varepsilon_{ij}^0 = Q_{ijkl}P_kP_l \quad \textbf{Eq. 11}$$

where $Q_{ijkl}$ is the electrostrictive tensor. The total strain in the system is calculated as:

$$\varepsilon_{ij} = \varepsilon_{ij}^{s} + \delta\varepsilon_{ij} \quad \textbf{Eq.12}$$

where the homogeneous strain ($\varepsilon_{ij}^{s}$) is induced by substrate clamping and the heterogeneous strain is related to the mechanical displacement field through the definition:

$$\delta\varepsilon_{ij} = \frac{1}{2}\left(\frac{\partial u_i}{\partial x_j} + \frac{\partial u_j}{\partial x_i}\right) \quad \textbf{Eq. 13}$$

The displacement field is determined by solving mechanical equilibrium equation:

$$\frac{C_{ijkl}(\partial^2 u_k)}{\partial x_j \partial x_l} = \frac{C_{ijkl}\left(\partial \varepsilon_{kl}^{0}\right)}{\partial x_j} \quad \textbf{Eq. 14}$$

The bottom of the film was mechanically clamped ($u_x$=$u_y$=0) via Dirichlet boundary conditions to simulate a rigid substrate, while the top surface was stress-free. The elastic constants of the amorphous $Al_2O_3$ interlayer are $C_{11}$=381.4 GPa, $C_{12}$=99.4 GPa, and $C_{44}$=141.0 GPa,[76] while the constants for HZO are $C_{11}$=450.1 GPa, $C_{12}$=124.0 GPa, $C_{44}$=5.47 GPa, $Q_{11}$ = 0.030 $m^4C^{-2}$ and $Q_{12}$ = −0.015 $m^4C^{-2}$.[16] At each time step, Poisson's equation was solved to account for changes in electrostatic potential fields, where V is the electrostatic potential across the model. The electrostatic potential was applied using Dirichlet boundary conditions at the top and bottom of the model. The electrostatic equilibrium of the system was defined as:

$$\frac{\partial}{\partial x_j}\left(\varepsilon_r(r)\varepsilon_0 \frac{\partial V(r)}{\partial x_j}\right) = \frac{\partial P}{\partial x_j} - \rho(r) \quad \textbf{Eq. 15}$$

where $\varepsilon_r$ is the relative permittivity, $\varepsilon_0$ is the vacuum permittivity (8.85 x $10^{-12}$ F/m), and $\rho$ is the electric charge density.

*Charge Transfer Mechanism*

To model the oxygen vacancy occupancy dynamics and electron transfer across the dielectric interface, a tunneling mechanism was adapted from Fontanini et al.[49] and Alhada-Lahbabi et al.[16] While other charge transfer mechanisms are likely to occur in these dielectric devices with the thickness of the dielectric material (Poole-Frenkel, Schottky emission, etc.), the models focus on tunneling as the primary means of charge transfer, which is supported by further thickness dependent P(E) loops and consistent with reports of the impact of tunneling in HZO devices in prior literature.[15]

Oxygen vacancy traps are assumed to form with an equal concentration on the top and bottom of the interface, with an equal charge concentration of positively charged oxygen vacancies ($V_O^{\cdot\cdot}$) and electrons ($e^-$). Due to the different timescales of electronic and ionic transport, the oxygen vacancies are assumed to be static while the electron concentration was allowed to evolve with time. The total electron concentration was held constant between the top and bottom interfaces. The evolution of trapped electrons on either side of the interface is given by a bidirectional tunneling model based on the Wentzel-Kramers-Brillouin (WKB) approximation for a trapezoidal barrier.[77] The effective barrier height for electron tunneling from the trap state across the dielectric is given by:

$$\phi_b = E_{trap} + \Delta E_{CB} \qquad \textbf{Eq. 16}$$

where $E_{trap}$ is the trap depth below the ferroelectric conduction band and $\Delta E_{CB}$ is the conduction band offset between the ferroelectric and dielectric. Under the applied electric field, the barrier becomes trapezoidal with entry height $\phi_0 = \phi_b$ and exit height:

$$\phi_1 = \phi_b - q * E * t_d \qquad \textbf{Eq. 17}$$

where q is the elementary charge, E is the electric field ($E = -\frac{\partial V(r)}{\partial x_j}$), and $t_d$ is the dielectric layer thickness. The tunneling transmission coefficient, $T_t$, was calculated by:[78]

$$T_t = e^{-\frac{4\sqrt{2m*_e}}{3\,\hbar\,q\,E_d}*(\phi_0^{\frac{3}{2}} - \phi_1^{\frac{3}{2}})} \qquad \textbf{Eq. 18}$$

where $m^*_e$ is the effective tunneling mass of an electron (0.25 * $m_e$),[49] ħ is the reduced Planck constant, and $E_d$ is the electric field across the dielectric layer. When the voltage drop across the dielectric exceeds the barrier height, the tunnel barrier can be approximated by the standard Fowler-Nordheim expression,[78]

$$T_t = e^{-\frac{4\sqrt{2m*_e}}{3\,\hbar\,q\,E_d}*(\phi_0^{\frac{3}{2}})}$$ **Eq. 19**

The rate of charge transfer is then calculated as:

$$\Gamma_{b\to t} = \frac{\sigma_E \sigma_T m *_e}{2\pi^2\hbar^3} k_b T * T_t * n_b * (1 - \frac{n_t}{n_{max}})$$ **Eq. 20**

where, $\Gamma_{b\to t}$ is the transfer rate from the bottom to the top interface, $\sigma_E$ is the energy cross-section (7 x $10^{-3}$ eV), and $\sigma_T$ is the trap area ($10^{-16}$ $cm^2$).[16,49] The energy cross-section and trap area values used for the model are from ref [49]. $k_b$ is Boltzmann's constant, T is temperature, set to 300 K, $n_b$ ($n_t$) is the number of filled trap states at the bottom (top) HZO-$Al_2O_3$ interface (0.6 $nm^{-2}$). The transfer rate is approximated by evaluating the transmission coefficient with a single energy distribution. Likewise, the transfer rate from the top interface to the bottom interface was set to:

$$\Gamma_{t\to b} = \frac{\sigma_E \sigma_T m *_e}{2\pi^2\hbar^3} k_b T * T_t * n_t * (1 - \frac{n_b}{n_{max}})$$ **Eq. 21**

The charge trapping equations were solved at each timestep along with the time-dependent Landau-Ginzburg-Devonshire equation, elastic equilibrium, and Poisson's equation. The coupled system of equations was solved using the Newton-Raphson method with a first-order backward Euler time integration.[79] Convergence at each timestep was achieved when the nonlinear residual satisfied a relative tolerance of $10^{-8}$ and an absolute tolerance of $10^{-10}$.

*Kelvin Probe Force Miscroscopy*

KPFM was carried out on a Vero AFM (Oxford Instruments Asylum Research) in ambient conditions, using the interferometric displacement sensor (IDS) for imaging. Measurements used BudgetSensors Multi75E-G AFM probes (Cr/Pt coated, spring constant ~3 N $m^{-1}$). Charge was written in a single contact-mode pass over a 3 x 3 μm field, with the applied tip bias recorded at every pixel: a central 1 $\mu m^2$ square at +1.0 V within a surrounding region at -1.0 V, relative to the grounded bottom electrode. Charge of the same sign as the applied bias is left at the surface: at +1 V, electrons are drawn from the sample into the tip, equivalently holes are injected, and the resulting shift in local surface potential appears in the KPFM nulling bias. At ±1 V across the ~19 nm films the mean field (~0.5 MV $cm^{-1}$) is below the coercive field, so the resulting contrast reflects charge injection and trapping. The same area was then imaged continuously in lift mode (50 nm lift height) over 5 x 5 μm at 64 x 64 pixels, one frame every 64 seconds, beginning immediately after writing. Each written region was delineated from the recorded write bias area and the mean surface potential was referenced to the unwritten area of the same frame to remove probe drift. The resulting decay was fitted to a stretched-exponential:

$$\Delta V(t) = A * e^{-(\frac{t}{\tau})^{\beta}} + C \quad \textbf{Eq. 22}$$

where A is the decaying amplitude, $\tau$ is the characteristic relaxation time, and $\beta$ is the stretching exponent, which describes charge release from a dispersive and inhomogeneous trapping landscape.

*Simulated First Order Reversal Curves*

The first order reversal curves (FORC) were simulated using the same phase-field procedure described above. To accurately capture the transient response, a dynamic kinetic coefficient that depends on the rate of change in polarization was used wherever partial switching is significant, as described in [80]. For the decreasing field, the effective kinetic coefficient was 2.9 x $10^{-6}$ $\Omega^{-1}cm^{-1}$, the same as used in the full switching loops. For the increasing field in the partial switching regime, the coefficient was set to 7.9 x $10^{-5}$ $\Omega^{-1}cm^{-1}$. The kinetic coefficient after the partial switching regime for increasing field reverted to 2.9 x $10^{-6}$ $\Omega^{-1}cm^{-1}$. Partial switching was found to occur up to 0.05 MV/cm for the 19H and 0.22 MV/cm for 8H-3A-8H devices. The contour plot overlaid on the experimental FORC data represent switching density > 80% of the highest simulated switching density.

# Supplemental Information

# Tuning the Coercive Field in Ferroelectric $Hf_{0.5}Zr_{0.5}O_2$-$Al_2O_3$ Heterostructures via Interfacial Charge Dynamics

Marshall Frye[1], Chanyoung Kim[1], Jeong-Woo Sun[1], John Wellington-Johnson[1], Lance Fernandes[2], Prasanna Venkatesan Ravindran[2], Bogdan Dryzhakov[3], TaeYoung Song[2], Mengkun Tian[4], Asif Khan[1,2], Lauren M. Garten[1]*

[1] School of Materials Science and Engineering, Georgia Institute of Technology, Atlanta, Georgia 30332, United States

[2] School of Electrical and Computer Engineering, Georgia Institute of Technology, Atlanta, Georgia 30332, United States

[3] Center for Nanophase Materials Sciences, Oak Ridge National Laboratory, Oak Ridge, Tennessee 37831, United States

[4]The Institute for Matter and Systems, Georgia Institute of Technology, Atlanta, Georgia 30332, United States

Corresponding Author: *Lauren.garten@mse.gatech.edu

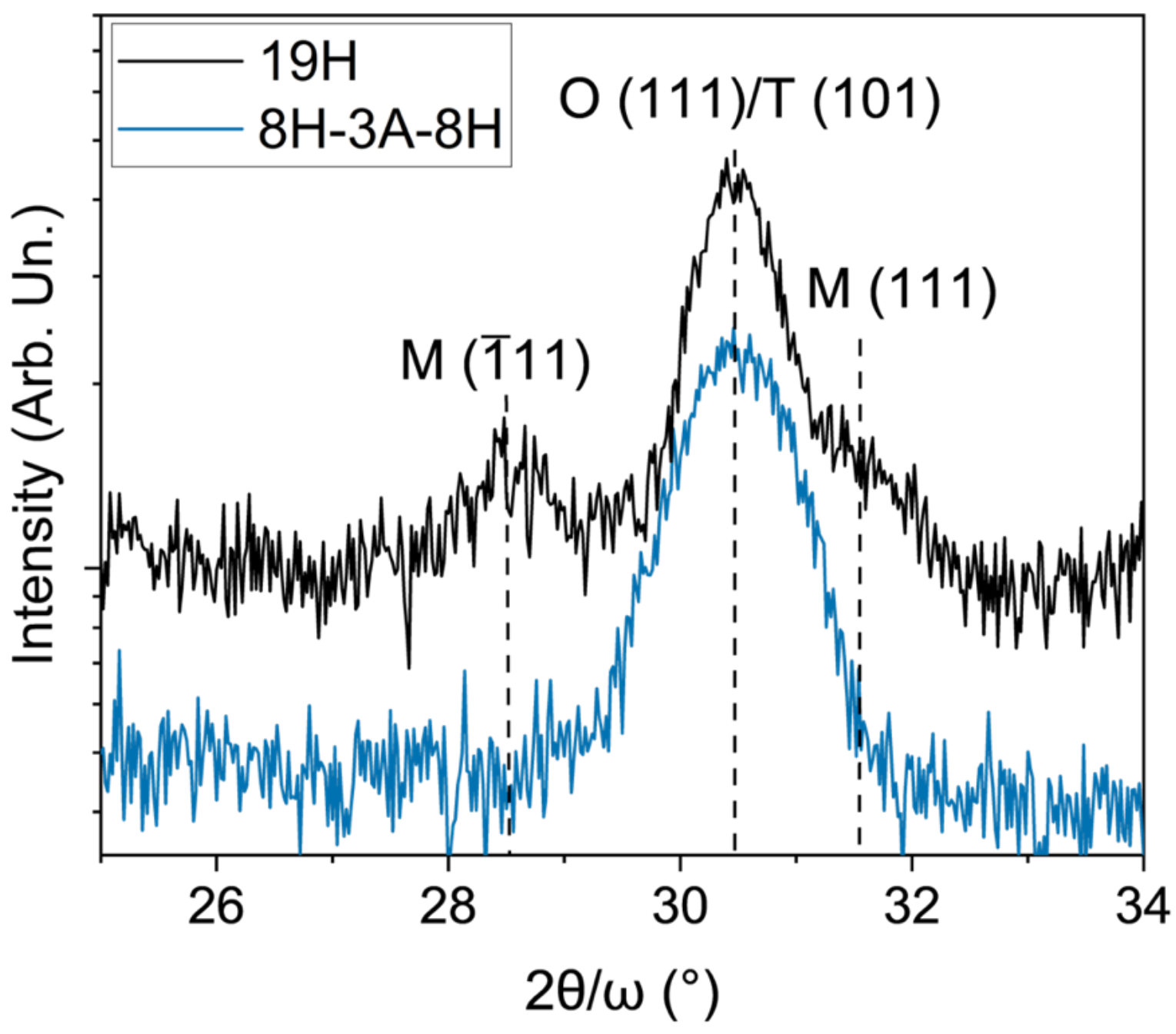

**Figure S1.** Grazing incidence X-ray diffraction of HZO samples with and without dielectric interlayers.

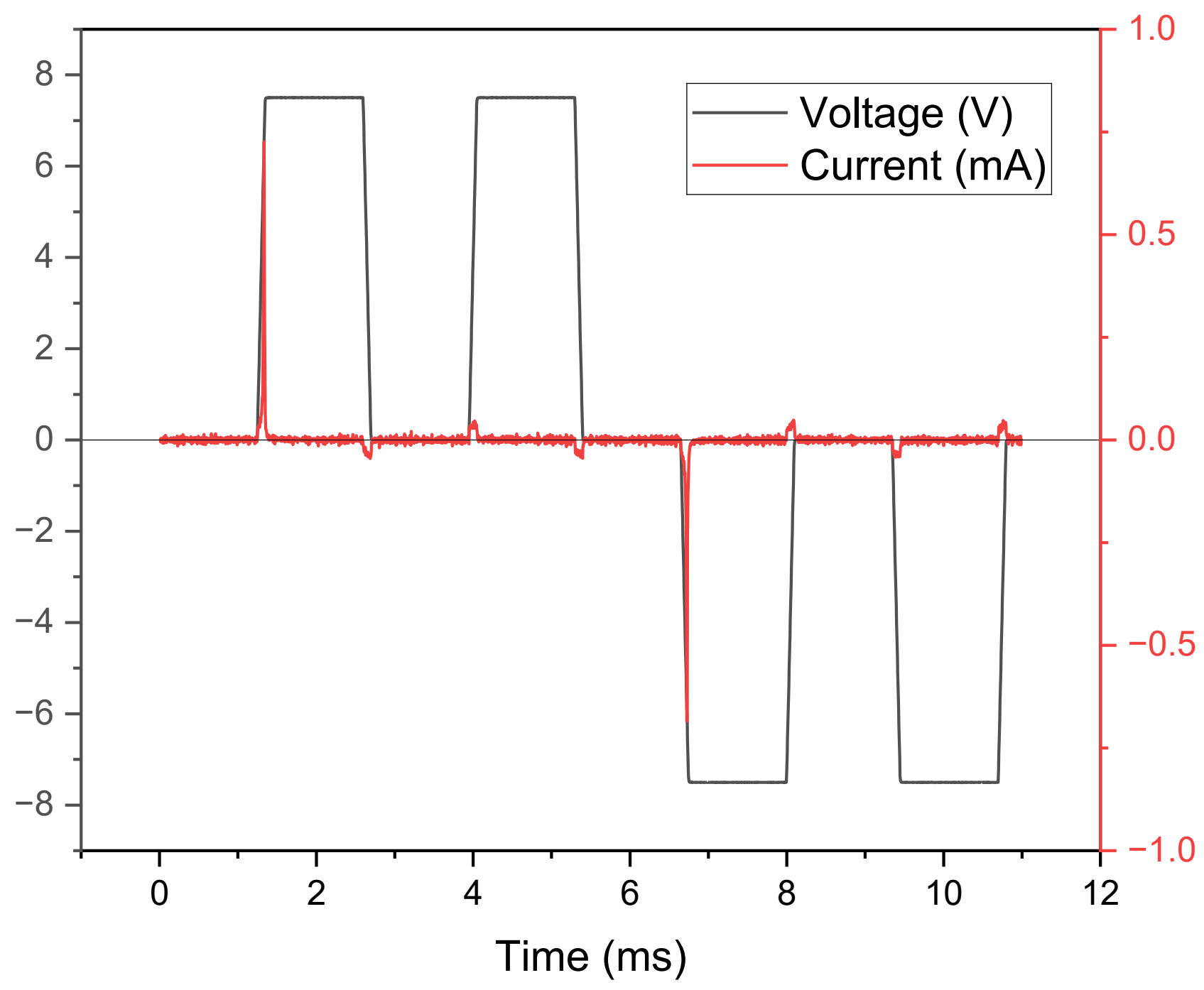

**Figure S2.** The voltage profile and resulting current for a Positive-Up-Negative-Down (PUND) measurement taken on an 8H-3A-8H device. The positive and negative PUND polarizations are 22.35 μC/cm$^2$ and 25.23 μC/cm$^2$, respectively.

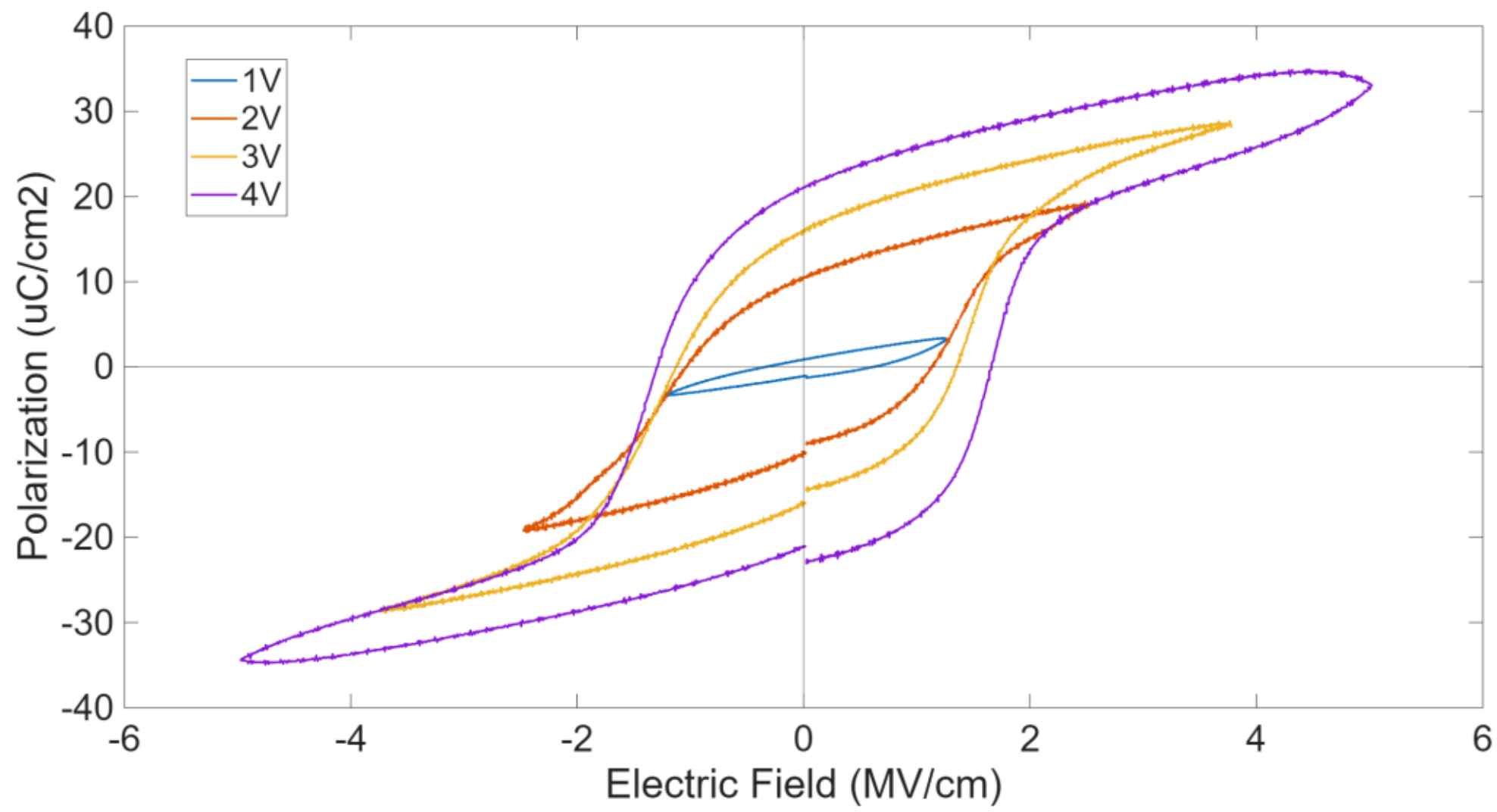

**Figure S3.** Polarization-electric field (PE) hysteresis loops of 8 nm continuous HZO device. The positive and negative coercive field values of the loop at 4V are 1.65 MV/cm and -1.29 MV/cm, respectively. The positive and negative remanent polarization values for the same loop are 21.05 μC/cm$^2$ and 22.16 μC/cm$^2$, respectively.

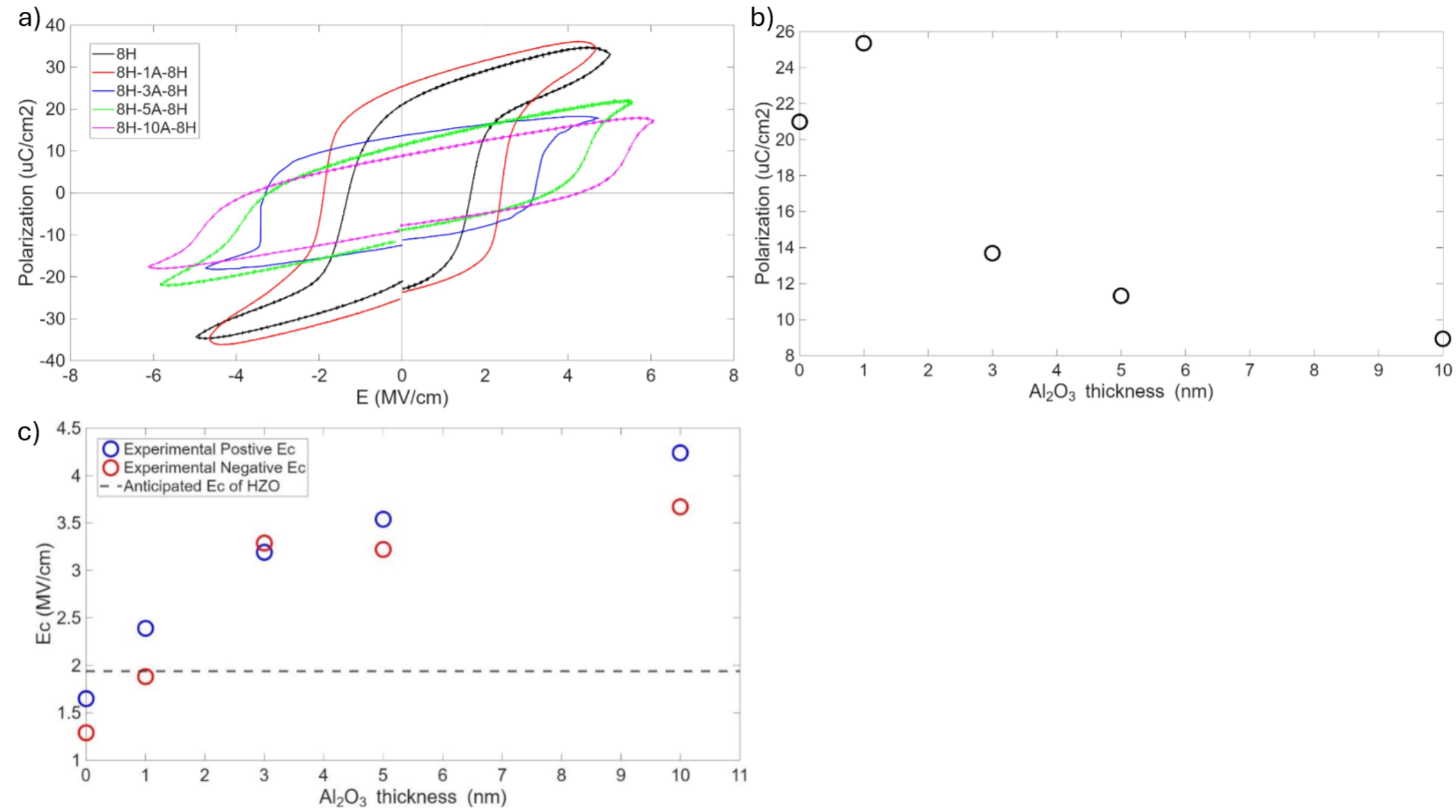

**Figure S4a.** PE hysteresis loops of HZO devices with varying $Al_2O_3$ dielectric interlayer thickness; **S4b.** Positive remanent polarization extracted from the PE data in S4a for devices with varied dielectric interlayer thickness; **S4c.** Positive and negative coercive fields extracted from PE data in S4a compared to theoretically estimated coercive field that would be anticipated across just the HZO layer.

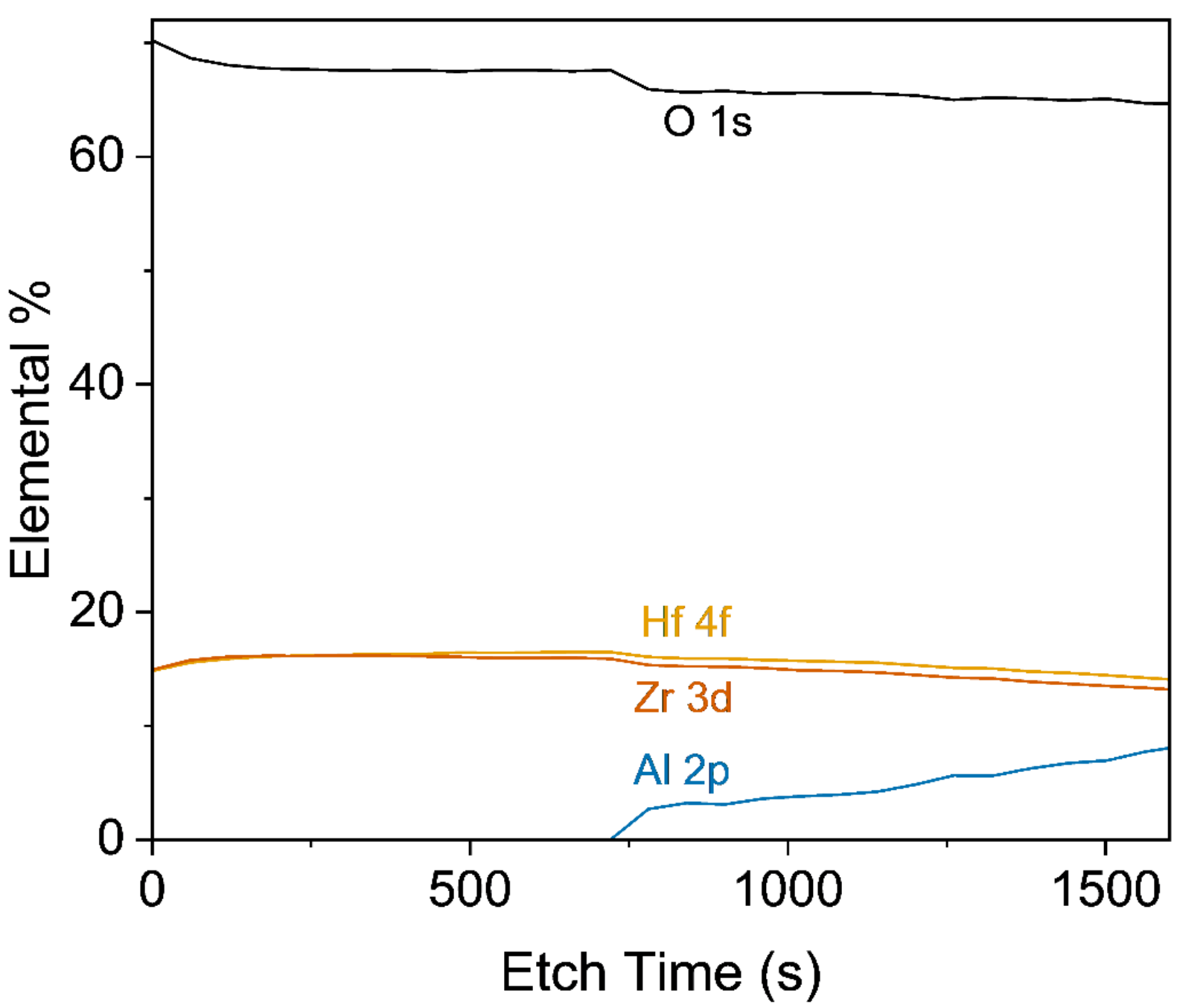

**Figure S5.** XPS depth profile a representative 8H-3A-8H sample. Zirconium and hafnium peaks are fit as a single oxide peak due to the reduced resolution of measurements taken during the depth profile.

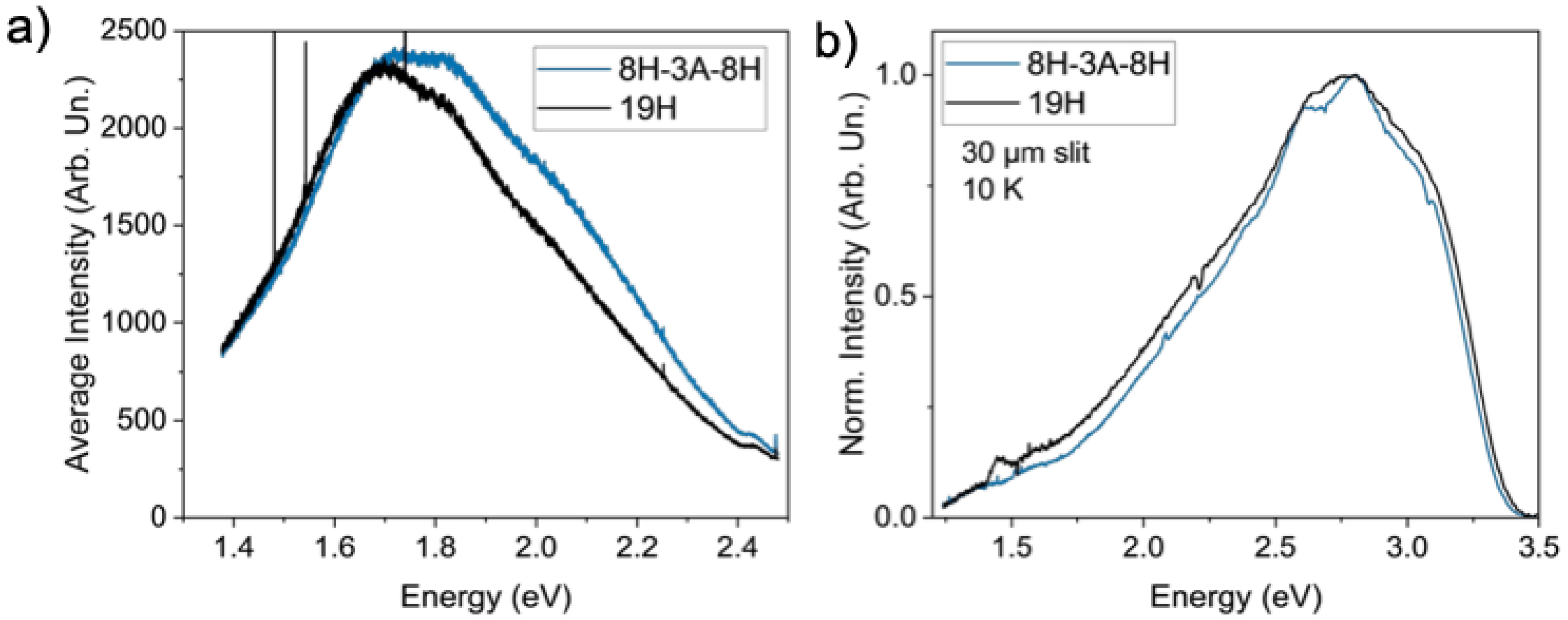

**Figure S6a.** Photoluminescence of the interlayer and control devices averaged across three random spots on each device; **S6b.** Cathodoluminescence of the interlayer and control devices measured at 10 K.

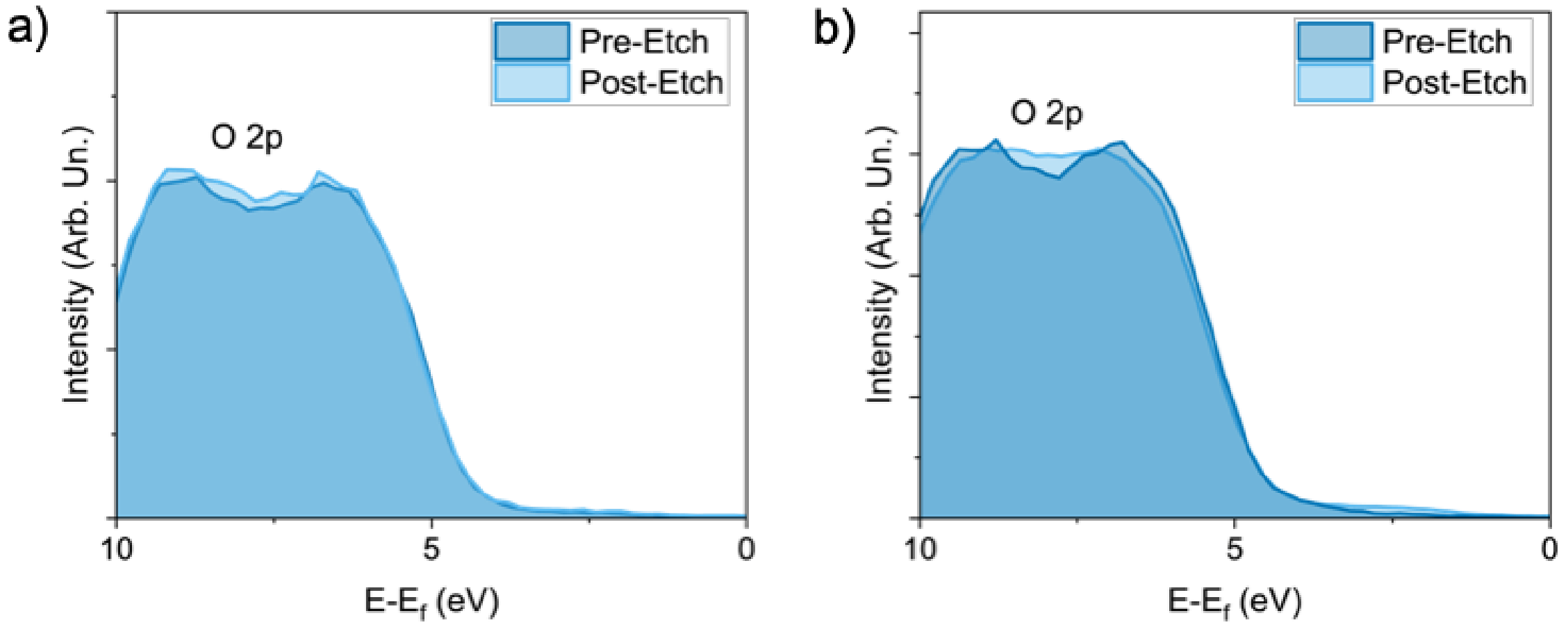

**Figure S7**. Valence spectra of the **a.** 19H and **b.** 8H-3A-8H devices before and after etching.

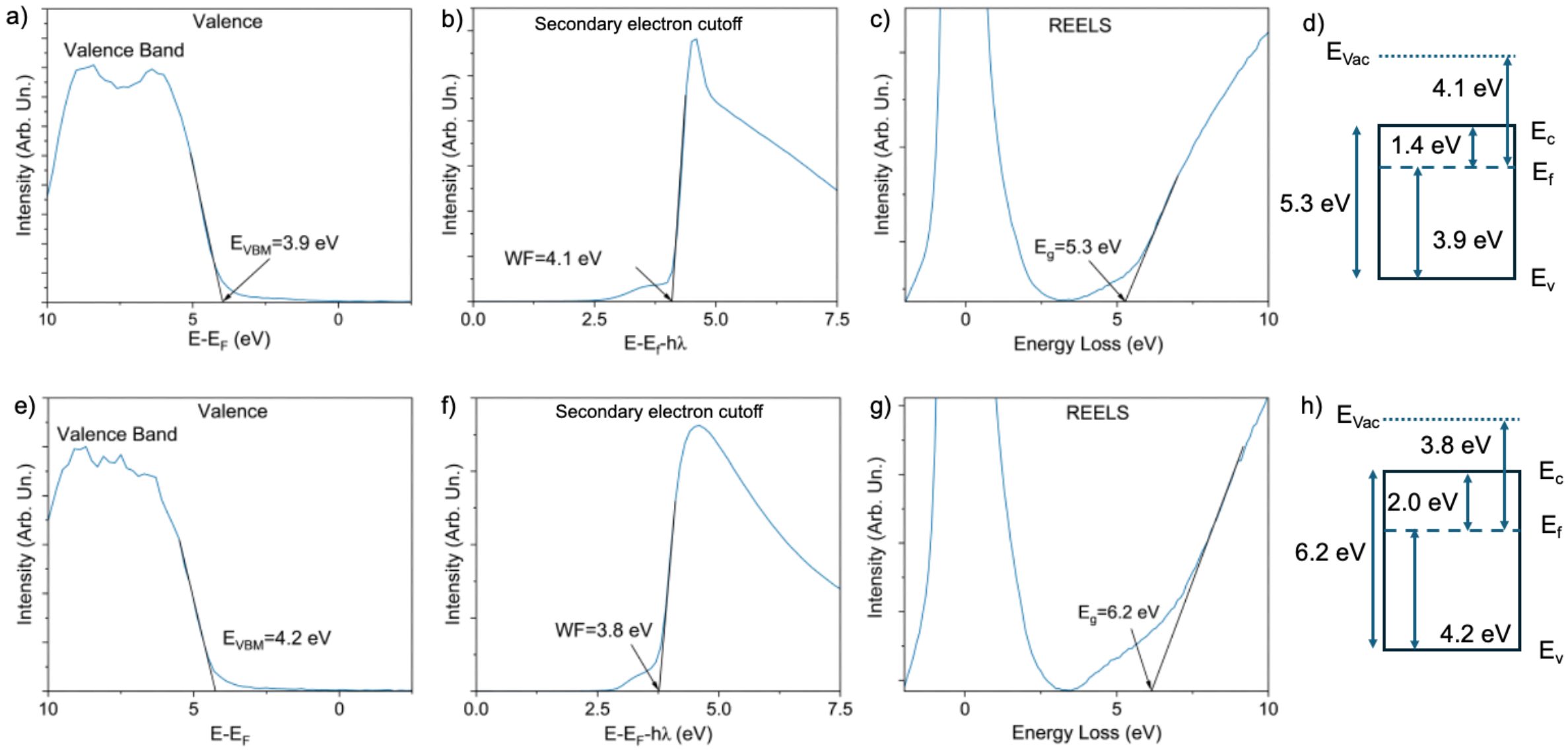

**Figure S8. a.** Valence spectra, **b.** secondary electron cutoff spectra, and **c.** REELS of HZO. **d.** Measured band gap of HZO. **e.** Valence spectra, **f.** secondary electron cutoff spectra, and **g.** REELS of $Al_2O_3$. **h.** Measured band diagram of $Al_2O_3$.

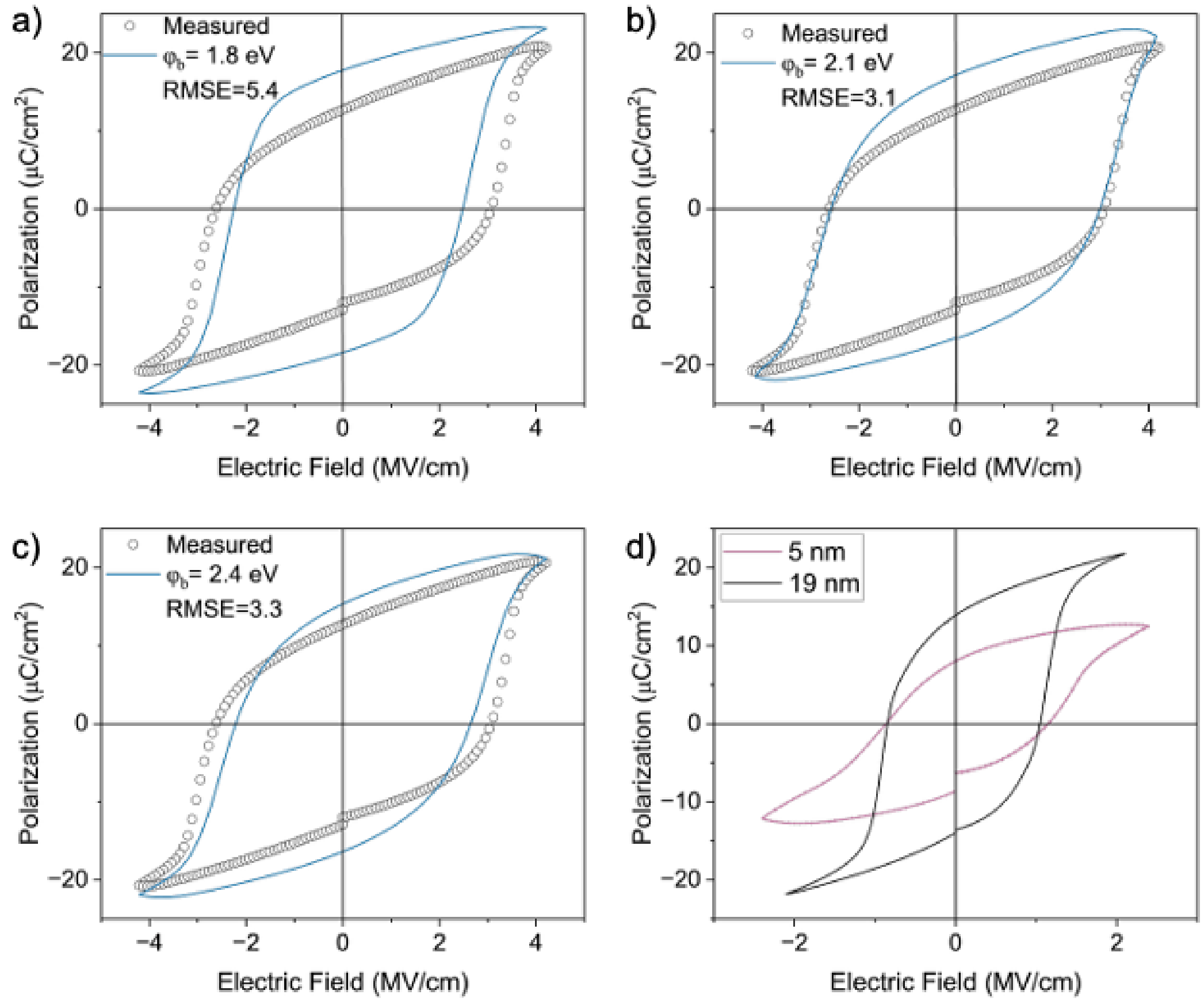

**Figure S9**. Phase field model of the interlayer 8H-3A-8H sample compared to representative experimental results as a function of tunnel barrier. The tunneling barrier was modified from **a.** 1.8 eV, corresponding to a defect energy of 1.2 eV, to **b.** 2.1 eV (corresponding to a defect energy of 1.5 eV) to **c.** 2.4 eV (corresponding to a defect energy of 1.8 eV). **d.** P(E) hysteresis of a 5 nm and 19 nm thick HZO film, showing comparable $E_c$ and reduced $P_r$ with reduced thickness.

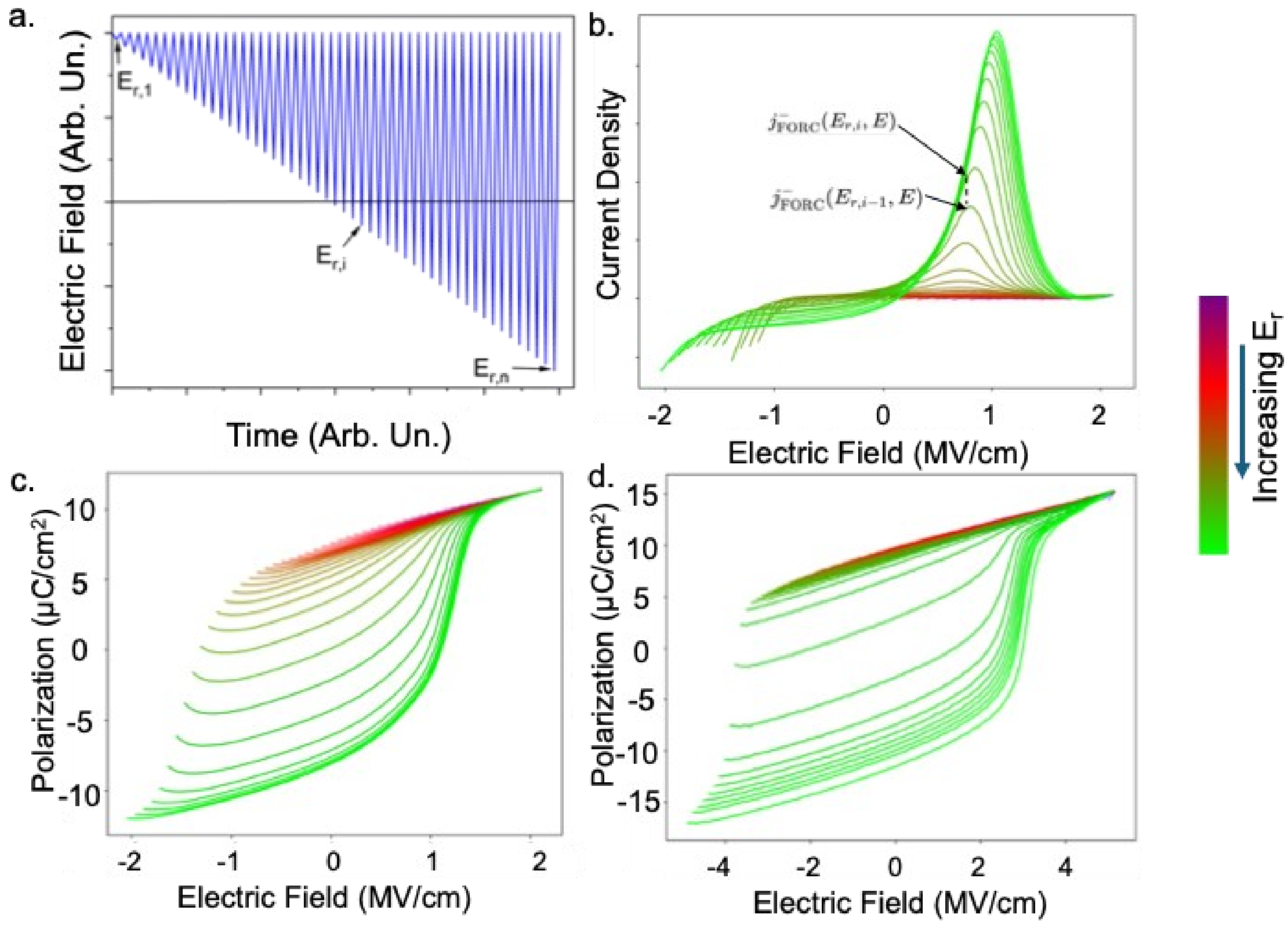

**Figure S10. a.** Waveform used to measure the FORC. **b.** Current density-electric field hysteresis measured from the 19H sample during the FORC measurement. The dashed line shows a region that is used to calculate the switching density between two different reversal curves. Polarization-electric field hysteresis with increasing reversal field in the **c.** 19H and **d.** 8H-3A-8H samples.

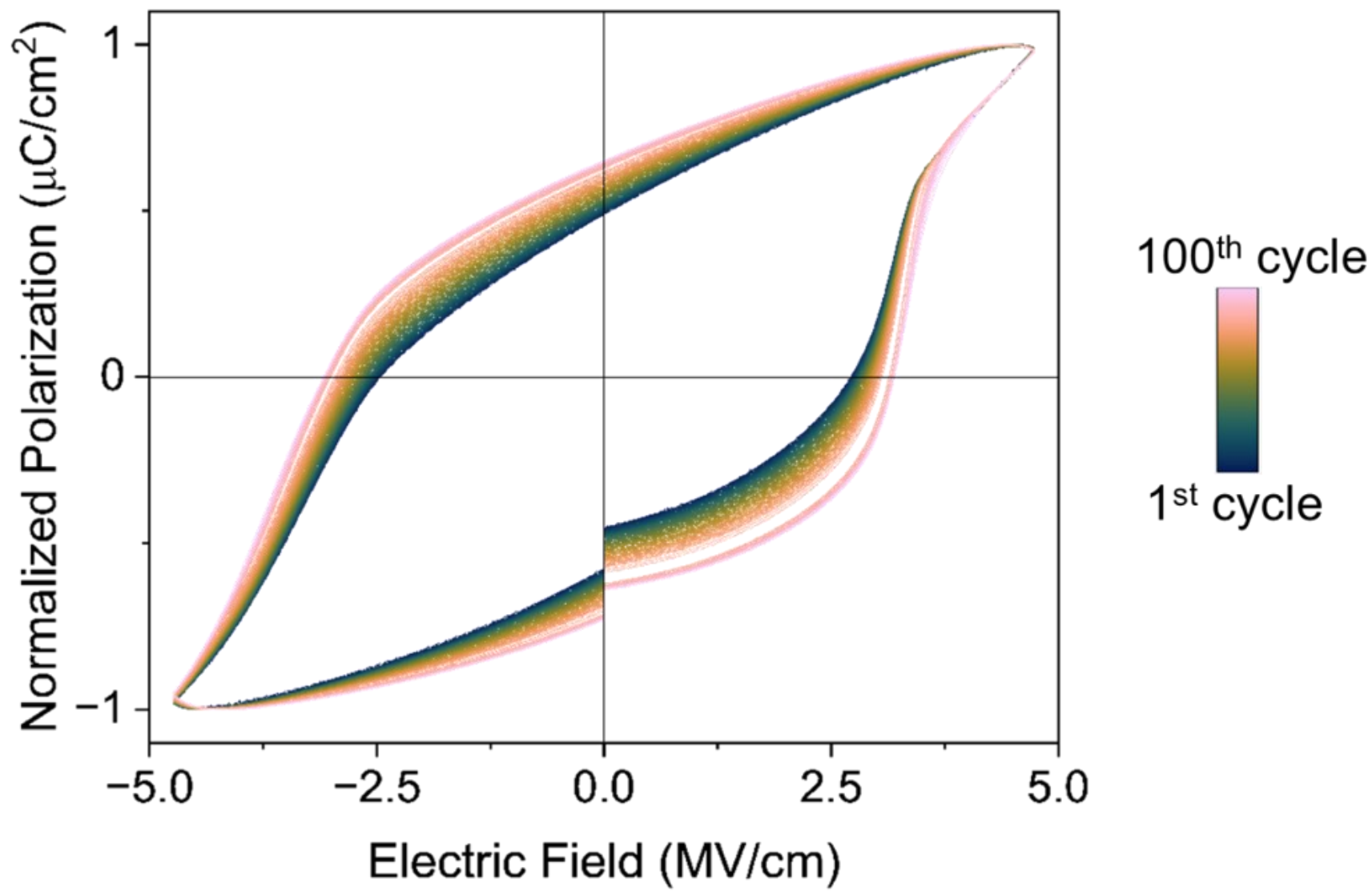

**Figure S11.** Cycling of an HZO sample with no top electrode using a removable indium contact.

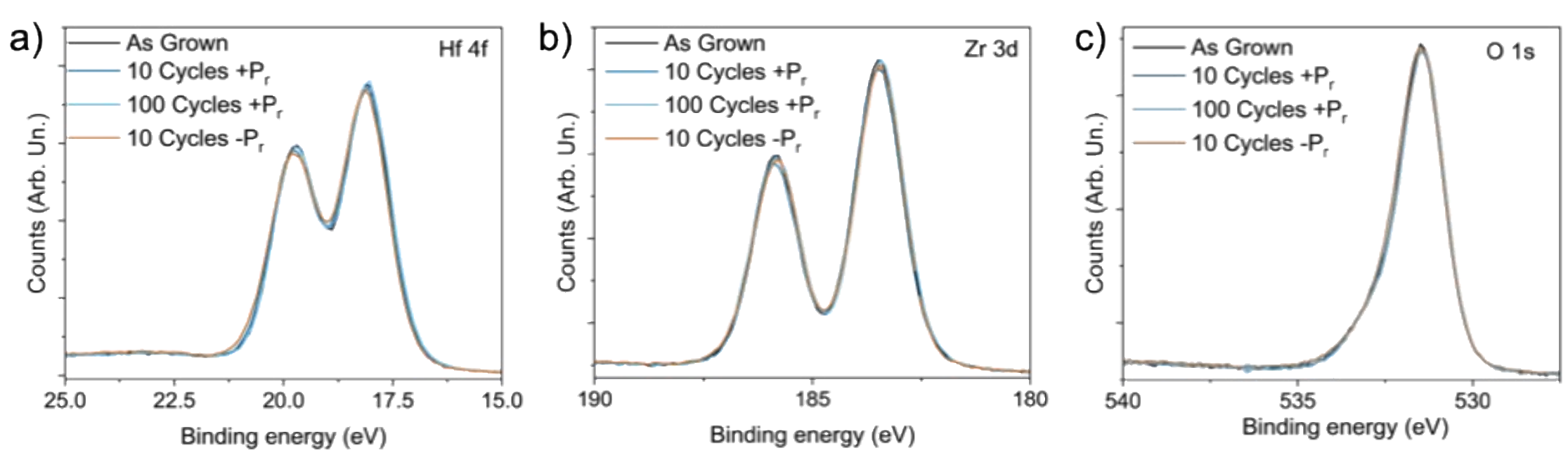

**Figure S12**. Polarization-dependent XPS core level spectra of **a.** Hf 4f, **b.** Zr 3d, and **c.** O 1s.

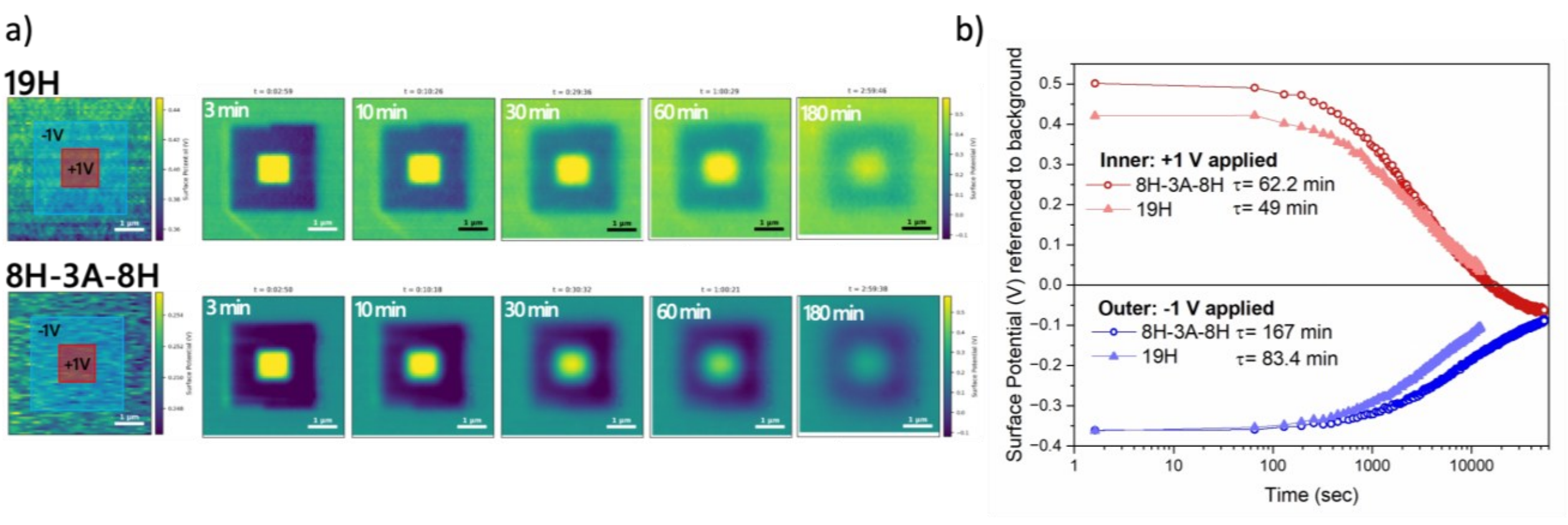

**Figure S13.** Charge retention in bulk and interlayer HZO measured by time-resolved KPFM. **a.** KPFM surface-potential maps of the 19 nm bulk film (19H, top row) and the 8/3/8 nm interlayer film (8H-3A-8H, bottom row). The leftmost panels show each surface before biasing, with the written pattern overlaid: a 3 x 3 μm region held at -1 V enclosing a central 1 x 1 μm square held at +1 V. After tip-bias writing, the same area was imaged roughly every minute for 14 hours (8H-3A-8H) and 3.5 hours (19H); representative panels are shown at 3, 10, 30, 60, and 180 min. **b.** Changes in measured surface potential serve as a proxy for induced charge and were tracked by masking the positively and negatively poled regions and referencing each to the unwritten surface. For each trace, the decay time constant (τ) was fit using a stretched exponential. The 8H-3A-8H sample shows a longer decay time constant than the 19H sample in both written regions (62.2 vs. 49 min for the +1 V region and 167 vs. 83.4 min for the -1 V region) indicating that the electrically induced charge state is more stable and relaxes more slowly in the interlayer structure.

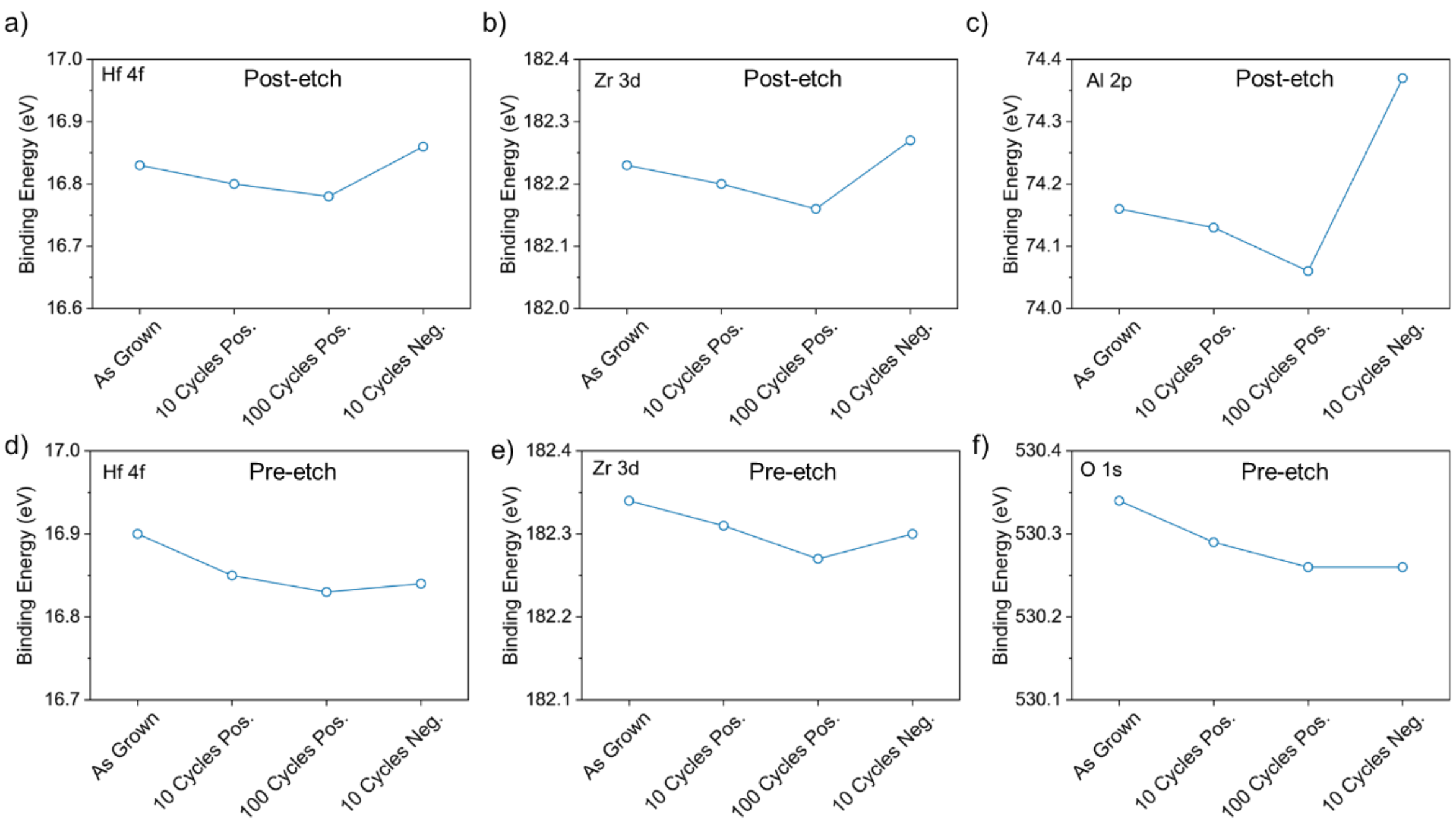

**Figure S14.** Binding energies of the **a.** $Hf^{4+}$, **b.** $Zr^{4+}$, and **c.** $Al^{3+}$ XPS peaks with cycling after etching. Binding energies of the **d**. $Hf^{4+}$, **e.** $Zr^{4+}$, and **f.** O 1s metal-oxide XPS peaks before etching. Binding energy are referenced to the tungsten metal peak in the W 4f spectra at 31.6 eV.[78,79]

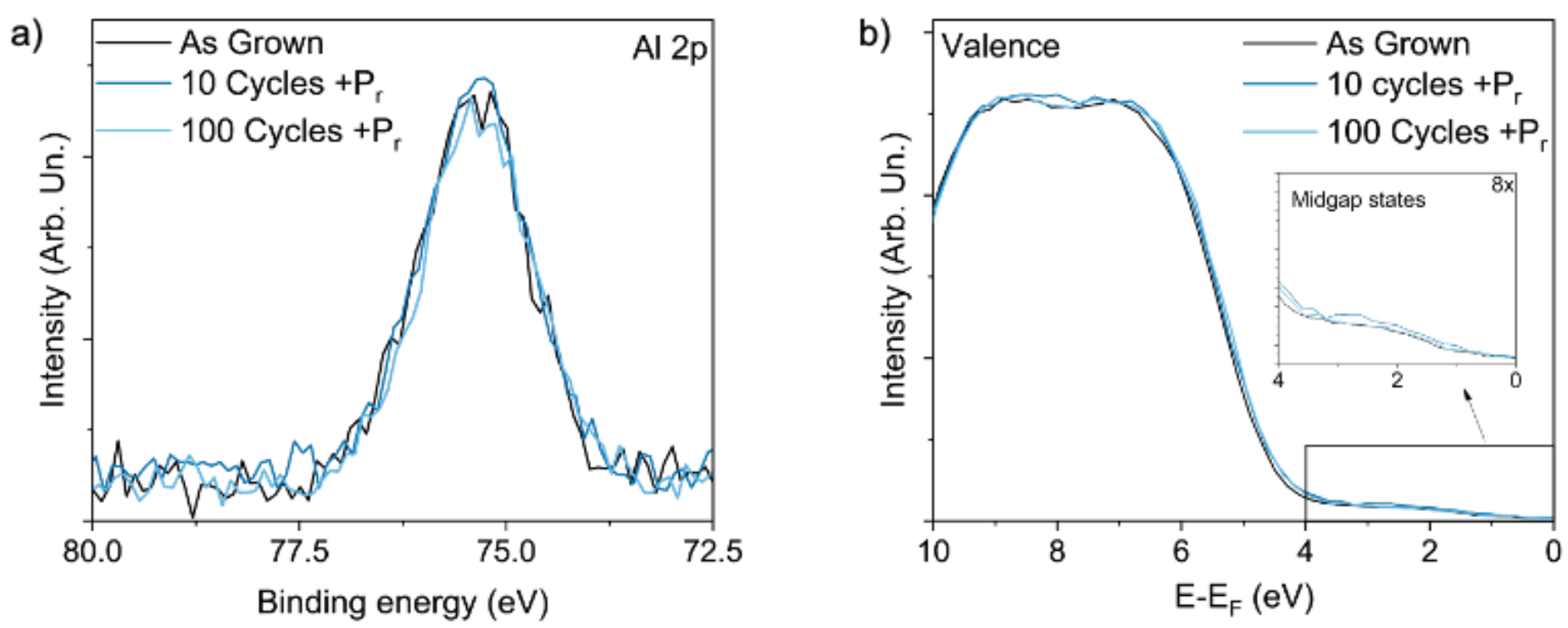

**Figure S15a.** Al 2p core-level and **S15b.** Valence XPS spectra as a function of cycling.

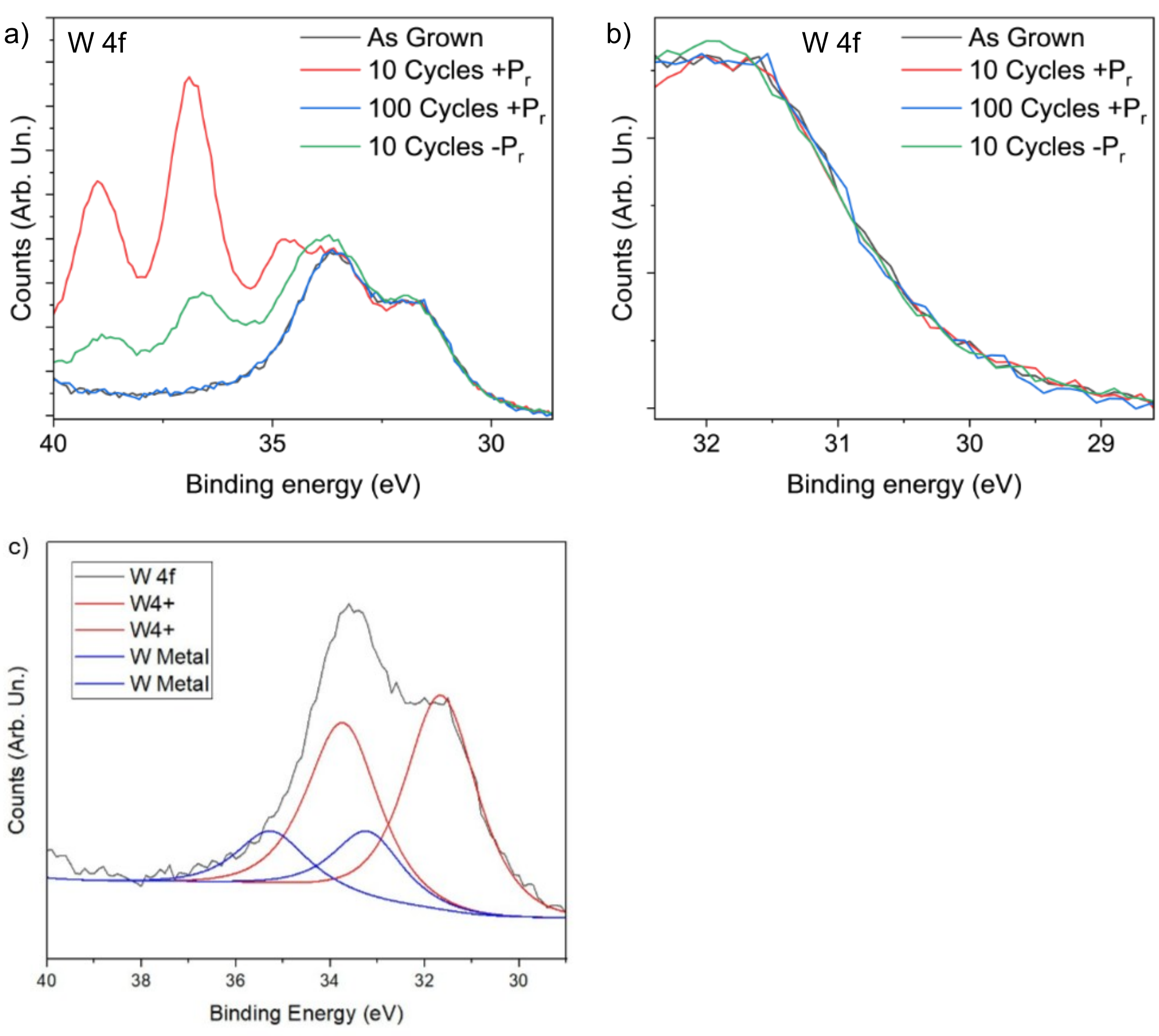

**Figure S16.** W 4f spectra of samples with poling. Peaks are charge shift referenced to the tungsten metal peak at 31.65 eV, which is within the range of previously reported values.[78,79] **S16a.** full W 4f spectra, showing that the tungsten on the devices show a range of oxidation state. **S16b.** Zoomed in W 4f spectra, showing that the trailing edge and metal peaks match after the charge-shift correction; **S16c.** Fitted W 4f spectra of the as grown sample showing the W metal peaks used as the charge shift reference and the 4+ suboxide peaks with the 4f7 peak positions at 31.65 eV and 33.19 eV, respectively.

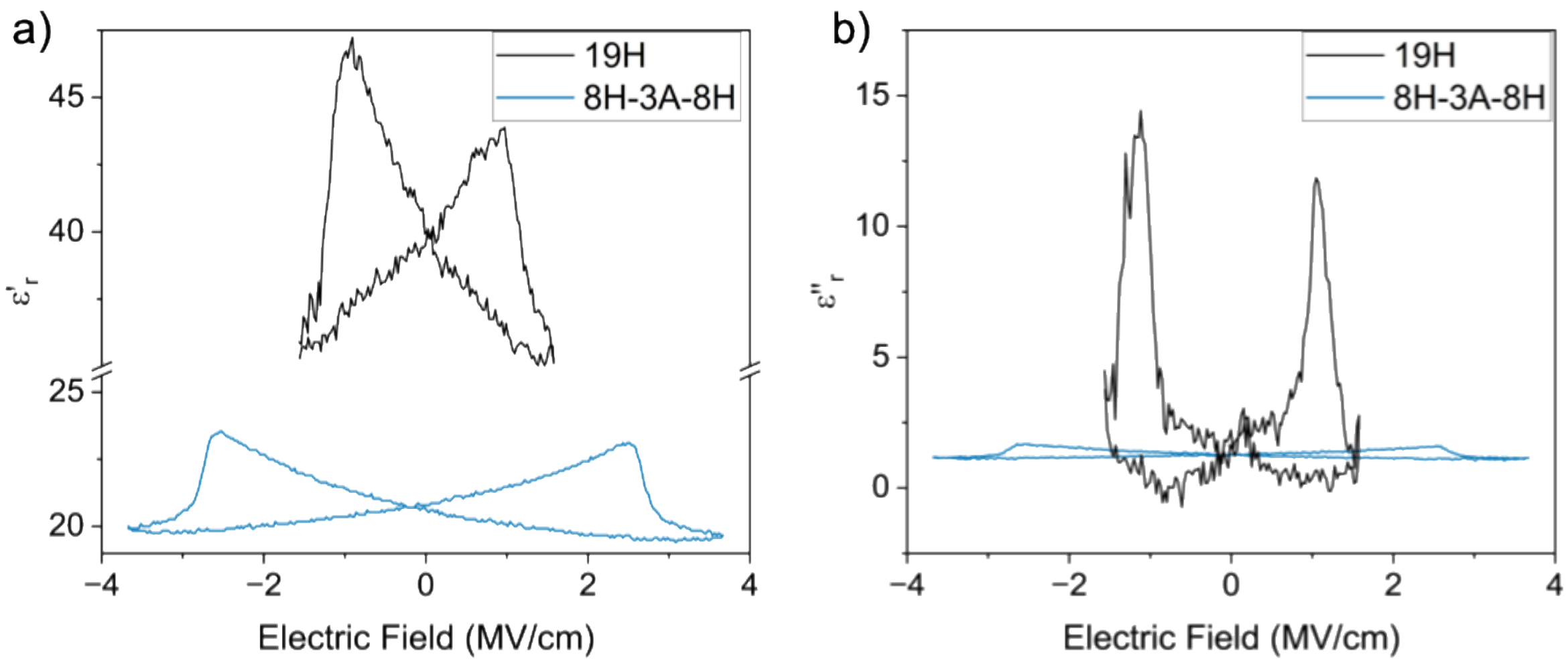

**Figure S17.** Capacitance-voltage hysteresis loop of HZO samples without (19H) and with (8H-3A-8H) an $Al_2O_3$ interlayer. **a,** real and **b,** imaginary components of the permittivity.

**Supplemental note 1**. Adding a dielectric material with a lower dielectric permittivity, such as $Al_2O_3$, would impact how the electric field drops across the ferroelectric layers, thus inducing an increase in the coercive field of the device. The first step in accessing the contributions that impact the coercive field is determining the field drop that is anticipated across each of the layers in the interlayer device. First, the permittivity is extracted from the capacitance of the control device. **Figure S17** shows the permittivity as a function of electric field, $\varepsilon_r(E_{DC})$, for each type of device. Again, both device types display the expected hysteresis for a ferroelectric material.[71] The measured relative permittivity of 19H control devices is 38.4, which is in the range of previously reported values for HZO deposited by ALD.[72] The total measured dielectric constant of the 8H-3A-8H is 21.1. Next the capacitance contribution of the $Al_2O_3$ layer is determined using the series capacitance rule,

$$\frac{1}{C_{total}} = \sum_i \frac{1}{C_i} \quad \textbf{Eq. 23}$$

where $C_i$ is the capacitance of each capacitor and $C_{total}$ is the equivalent capacitance of the circuit. From **Eq.** , the dielectric permittivity of the $Al_2O_3$ interlayer is 6.4, which is in the range of previously reported values ALD $Al_2O_3$ layers.[73,74] Then, to determine the electric field across each layer in the 8H-3A-8H device, the device was modelled as 3 parallel-plate capacitors in series (HZO-$Al_2O_3$-HZO). The voltage dropped across each capacitor is calculated by:

$$V_i = \frac{C_{total}}{C_i} * V_{total} \quad \textbf{Eq. 24}$$

where $V_i$ is the voltage dropped across a capacitor with capacitance $C_i$, $C_{total}$ is the equivalent capacitance of the circuit, and $V_{total}$ is the total voltage applied across the capacitor stack. Using **Eq.** , the calculated electric field across the HZO layers within the 8H-3A-8H interlayer device at the coercive field is 1.94 MV/cm. Even after accounting for the dielectric layer, there is still a significant increase in the coercive field with the insertion of the dielectric layer from the control devices where $E_c$ = 1.01 MV/cm. Therefore, while the $Al_2O_3$ layer decreases the effective electric field across the HZO, there is still a difference in the coercive field of the between different device architectures that cannot be accounted for by simply adding a material with lower dielectric permittivity in series with ferroelectric $Hf_{0.5}Zr_{0.}5O_2$. The disproportionate $E_c$ shows that there are additional contributions to the increase in coercive field.